\newcommand{\cratio}{$^{12}$C/$^{13}$C}
\newcommand{\nratio}{$^{14}$N/$^{15}$N}
\newcommand{\oratio}{$^{16}$O/$^{18}$O}

\documentclass[preprint2]{pp7}
\usepackage[version=4]{mhchem}
\usepackage{hyperref}
\usepackage[switch]{lineno}
\usepackage{color}
\newcommand{\rev}[1]{\textcolor{red}{\textsf{#1 }}}

\input{pp7.h}

\begin{document}


\title{\textbf{\LARGE The Isotopic {Links} from {Planet Forming Regions} to the Solar System}}

\author {\textbf{\large H. Nomura$^{1,2}$, K. Furuya$^1$, M.A. Cordiner$^{3,4}$, S.B. Charnley$^3$, C.M.O'D. Alexander$^5$, C.A. Nixon$^3$, V.V. Guzman$^6$, H. Yurimoto$^7$, T. Tsukagoshi$^1$, T. Iino$^8$}}
\affil{\small\it $^1$ Division of Science, National Astronomical Observatory of Japan, 2-21-1 Osawa, Mitaka, Tokyo 181-8588, Japan}
\affil{\small\it $^2$ The Graduate University for Advanced Studies, SOKENDAI, 2-21-1 Osawa, Mitaka, Tokyo 181-8588, Japan}
\affil{\small\it $^3$ Solar System Exploration Division, NASA Goddard Space Flight Center, Greenbelt, MD, USA}
\affil{\small\it $^4$ Department of Physics, Catholic University of America, Washington DC, USA}
\affil{\small\it $^5$ Earth and Planets Laboratory, Carnegie Institution of Washington, 5241 Broad Branch Road NW, Washington, DC 20015, USA}
\affil{\small\it $^6$ Instituto de Astrofísica, Pontificia Universidad Católica de Chile, Av. Vicuña Mackenna 4860, 7820436 Macul, Santiago, Chile}
\affil{\small\it $^7$ Department of Natural History Sciences, Hokkaido University, Sapporo, Hokkaido 060-0810, Japan}
\affil{\small\it $^8$ Information Technology Center, The University of Tokyo, 2-11-16, Yayoi, Bunkyo, Tokyo 113-8658, Japan}


\begin{abstract}
\baselineskip = 11pt
\leftskip = 1.5cm 
\rightskip = 1.5cm
\parindent=1pc
{\small 
Isotopic ratios provide a powerful tool for understanding the origins of materials, including the volatile and refractory matter within solar system bodies. Recent high sensitivity observations of molecular isotopologues, in particular with ALMA, have brought us new information on isotopic ratios of hydrogen, carbon, nitrogen and oxygen in star and planet forming regions as well as the solar system objects. Solar system exploration missions, such as Rosetta and Cassini, have given us further new insights. Meanwhile, the recent development of sophisticated models for isotope chemistry including detailed gas-phase and grain surface reaction network has made it possible to discuss how isotope fractionation 
in star and planet forming regions is imprinted into the icy mantles of dust grains, preserving a record of the initial isotopic state of solar system materials. This chapter reviews recent progress in observations of molecular isotopologues in extra-solar planet forming regions, prestellar/protostellar cores and protoplanetary disks, as well as objects in our solar system --- comets, meteorites, and planetary/satellite atmospheres --- and discusses their connection by means of isotope chemical models.
\\~\\~\\~}
\end{abstract}  


\section{\textbf{INTRODUCTION}}

\subsection{Solar system formation and isotope fractionation}

Isotopic ratios are used in various fields of research to trace the chemical evolution of different materials. They are valuable not only for studying the origins of objects in our solar system, but also for investigating chemical pathways in the terrestrial environment, biosphere, and living organisms. For relatively complex molecules, position-specific isotope analysis has been established as one of the useful methods to analyse the origin of materials, such as biosynthetic pathways. Isotope analysis of functional groups in organic material in meteorites has been a highly beneficial tool {for understanding} their structure and origin \citep[e.g.,][]{RN7566}. Even in the case of astronomical observations, anomalies of the $^{12}$C/$^{13}$C ratio at different positions of carbon-chain molecules have been found in molecular clouds \citep[e.g.,][]{takano98,sakai10,giesen20}. 

The formation of the solar system must be considered in the context of the formation of low-mass stars and their surrounding planetary systems {\citep[e.g.,][]{hayashi85,pineda22,tsukamoto22,drazkowska22}}. 
The composition of gas giant planets will be similar to that of nebula gas, while rocky/ice planets and small bodies reflect the composition of refractory material and ice in/on the dust grains from which they accreted. Inside solar system objects, isotopic fractionations are further enhanced/diminished by local processes.

In present-day molecular clouds, about 99\% of the baryonic mass exists in the gas phase, and the remaining 1\% comprises of dust grains. Once these clouds become dense enough for dust grains in the clouds to shield the interstellar ultraviolet (UV) radiation, a complex chemistry can proceed \citep[e.g.,][]{herbst09}. When the density becomes high enough and the temperature is low enough, gas-phase species begin to freeze out onto dust grain surfaces to form ice mantles. Consequently, in such regions, {isotopic fractionation that occurred in the gas phase is imprinted in the ices} \citep[e.g.,][]{brown89,yurimoto04}. Solid-state (grain surface) reactions between molecules in the ice can further enhance or reduce the degree of isotopic fractionation. Meanwhile, desorption of molecules from the ice mantles can modify the isotope fractionation of the gas phase.

Therefore, in order to understand the isotopic fractionation in ice, and thus determine the initial conditions of small, icy bodies and {planetesimals} in the solar system, we need to treat the chemical interactions between gas and ice mantles on dust grains. Recently, the development of sophisticated numerical modelling considering detailed gas-phase chemical reaction networks, with grain surface reactions, has improved our understanding of isotope chemistry in interstellar gas and ice. 
Together with advances in sensitivity of 
molecular isotopologue observations in star and planet forming regions in the era of ALMA, 
our ability to reconstruct the isotopic history of matter in the Galaxy has improved dramatically. 
Spacecraft missions to various solar system bodies, such as Stardust and {Rosetta} for comets {81P/Wild and 67P/Churyumov-Gerasimenko (hereafter 67P)}, Cassini in the Saturnian system, and Hayabusa (1 and 2) towards asteroids Itokawa and Ryugu, have brought us further information on 
the isotopic compositions of primitive solar system materials. Such theoretical and observational progress has made it possible to elucidate the chemical link 
through the history of isotopes, giving us new insights into the origins of the materials found in our solar system today.

\subsection{Isotope ratios in star and planet forming regions and the solar system}

Figure 1 summarizes the measurements of isotopic ratios D$/$H, \ce{^{12}C}$/$\ce{^{13}C}, \ce{^{14}N}$/$\ce{^{15}N}, and \ce{^{16}O}$/$\ce{^{18}O} in different molecules, for various objects from prestellar/protostellar cores to protoplanetary disks around low mass stars, comets, meteorites, and planets/satellites. \ce{^{16}O}$/$\ce{^{17}O} is not plotted due to the lack of measurements compared with other isotope ratios. {As mentioned, ice is formed on grain surfaces in cold and dense regions of prestellar cores. As soon as a star is formed and begins to heat the surrounding material, the dust temperature
exceeds the sublimation temperature of molecules in the ice close to the protostar, and the molecules are released into the gas} \citep[e.g.,][]{herbst09}. In order to compare with solar system objects, for some observations of protostellar objects (labelled `warm gas' in the figure), molecular species which are expected to be sublimated from ice are selected. Radial distributions of isotope ratios have been observed towards some protoplanetary disks (see Section 3 for details), but the averaged values are plotted in the figure. 

The deuterium-to-hydrogen (D/H) ratio in water is of particular interest in the context of the origin of water on Earth and other terrestrial-type bodies. Despite significant diversity, the D/H ratios of water in  comets and the bulk of meteorites are quite similar to that of terrestrial ocean water (see Section 4 and 6 for more details), in particular, with respect to the Vienna Standard Mean Ocean Water (VSMOW) value of $1.56\times10^{-4}$. 
The water D/H ratio in protostellar cores in clustered star-forming regions is similar to the high end of the water D/H ratio in comets \citep[e.g.,][]{persson14}, while the water D/H ratio in isolated protostellar cores is higher than that in comets \citep{jensen19,jensen21}.
This may be consistent with the scenario that the Sun was born in clustered star-forming regions \citep{adams10},
assuming that little or no processing of water occurs between the protostellar stage 
and the formation of planetesimals and comets \citep{persson14,jensen19}.
Meanwhile, deuterium ratios of organics are known to have higher values in protostellar objects, protoplanetary disks and comets (see Section 2). We note that water is usually the dominant component of ice in molecular clouds, disks, and comets, and fractional abundances of organic species relative to water are often smaller by more than two orders of magnitude. The D/H ratios of \ce{H2} gas in the atmospheres of giant planets are close to the reference (Solar) value, as expected given their accretion from a similar reservoir of nebular gas. Larger D$/$H ratios of the more evolved atmospheres of the inner planets occur as a result of differing atmospheric escape rates for isotopes of different masses (see Section 5).

Nitrogen species related to \ce{N2} (e.g., \ce{N2H+}) are \ce{^{15}N}-poor while HCN is \ce{^{15}N}-rich in protostellar objects and protoplanetary disks, which is qualitatively consistent with the theoretical prediction caused by isotope-selective photodissociation (see Section 2 and 3). Solar system objects have a \ce{^{15}N}-rich value partly due to the inheritance of interstellar material, and partly due to local processes in the solar system  (see Sections 5 and 6 for details). The exception is the atmospheres of gas giants whose nitrogen isotope ratios are close to the Solar reference value, probably keeping the value of the nebular gas.

Carbon and oxygen isotope ratios {have relatively small variations} among the solar system objects while they have large variations among the prestellar cores, depending on the species (see below). However, it is known that oxygen isotopes in meteorites in the solar system have systematic (but mass-independent) variations, which are interpreted 
by isotope selective photodissociation in molecular clouds/protosolar nebula (see Section 6 for details). 

\begin{figure*}[h!]
\centering
\includegraphics[width=0.75\linewidth]{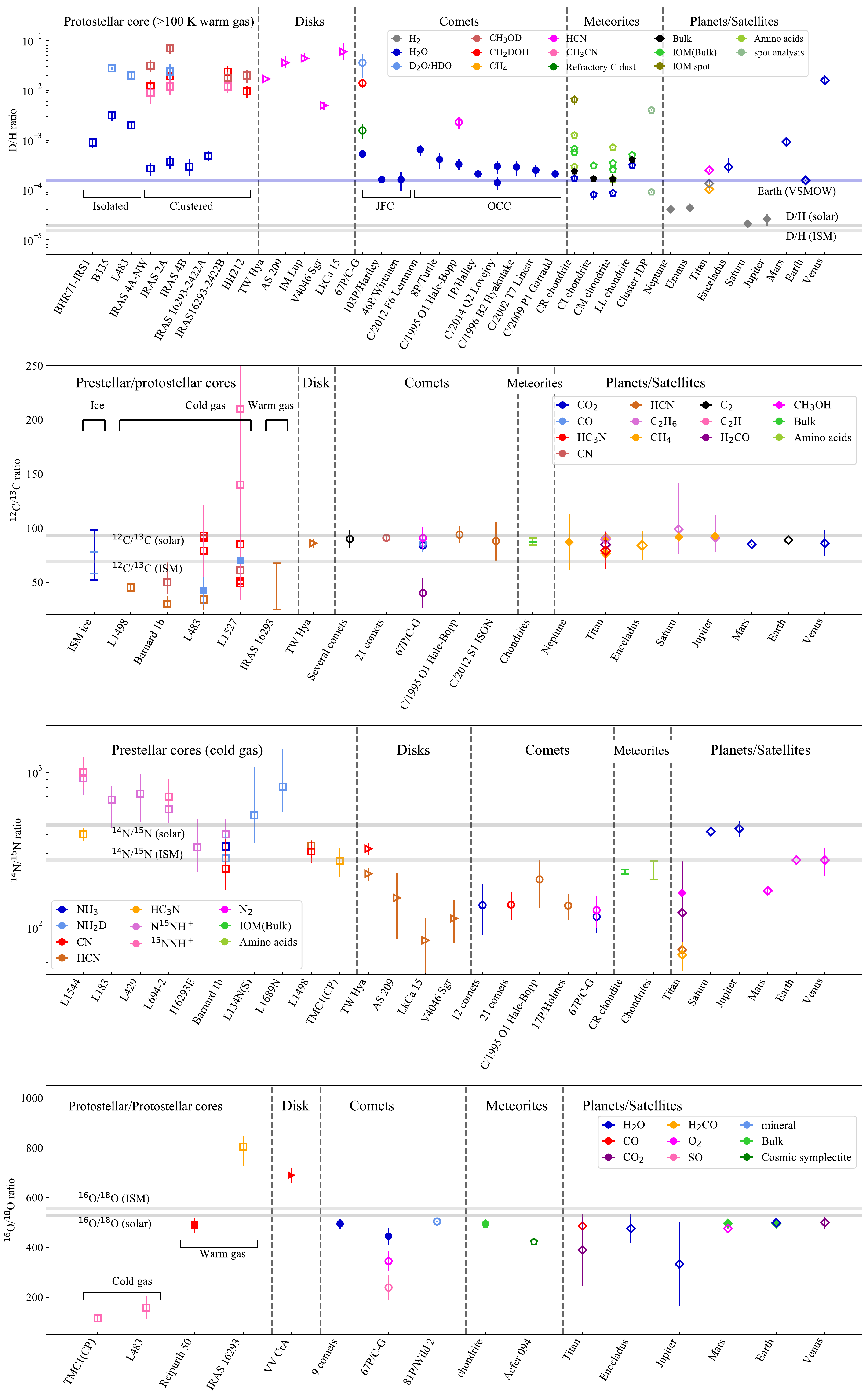}
\caption{Measurements of isotope ratios of D$/$H, \ce{^{12}C}$/$\ce{^{13}C}, \ce{^{14}N}$/$\ce{^{15}N}, and \ce{^{16}O}$/$\ce{^{18}O} for various objects, including prestellar/protostellar objects, protoplanetary disks, comets, meteorites, and planets/satellites. Different colors represent different molecules and different symbols show different types of object. Data points with error bars {($1\sigma$)} or ranges among multiple objects or various molecules are plotted.
Filled symbols show the dominant reservoir for a given element/object. Gray horizontal lines show reference (elemental) isotope ratios in the local ISM and in the solar system. Literature sources for these data are listed at \url{https://sci.nao.ac.jp/MEMBER/hnomura/ppvii_isotope/table.pdf}}
\label{fig:ratios}
\end{figure*}

We note that there are deviations in the reference isotope ratios between the local interstellar medium and the solar system. Elements heavier than hydrogen and helium, such as carbon, nitrogen, and oxygen, are produced in the interiors of stars through nuclear syntheses of hydrogen, which also determine their elemental isotope ratios \citep[e.g.,][]{romano19}. Therefore, the reference (elemental) isotope ratios evolve with the chemical evolution of the galaxies, and depend on the local star formation activity in the galaxies. The solar system was formed 4.5 billion years ago when heavier isotopes were less abundant than in the present-day local interstellar medium. Also, it is observationally known that the elemental isotope ratios have gradients inside our Galaxy depending on the distance from the Galactic Center, which can be reproduced by Galactic chemical evolution models and reflects the star formation activities in our Galaxy \citep[e.g.,][]{adande12,zhang15,colzi18,yan19}. Meanwhile, the observed reference (elemental) isotope ratios are not always consistent among different environments, for example, between diffuse and dense clouds. Measurements of isotope ratios are based on observations of molecular isotopologue lines, and we need to be careful about scatter of observed values caused by various processes/assumptions, such as fractionation due to chemical reactions (isotope selective photodissociation and exchange reactions), hyperfine anomalies \citep[e.g.,][]{daniel13,magalhaes18}, and excitation of transition lines \citep[e.g.,][]{hilyblant17}. Indirect measurements of isotope ratios, for example, deriving HCN/HC$^{15}$N ratios from H$^{13}$CN/HC$^{15}$N observations with assuming the \cratio\ ratios, also can introduce significant scatter \citep[][]{roueff15,hilyblant20}.


\section{\textbf{ISOTOPE FRACTIONATION CHEMISTRY}}

\begin{figure*}[t]
\centering
\includegraphics[width=0.75\linewidth]{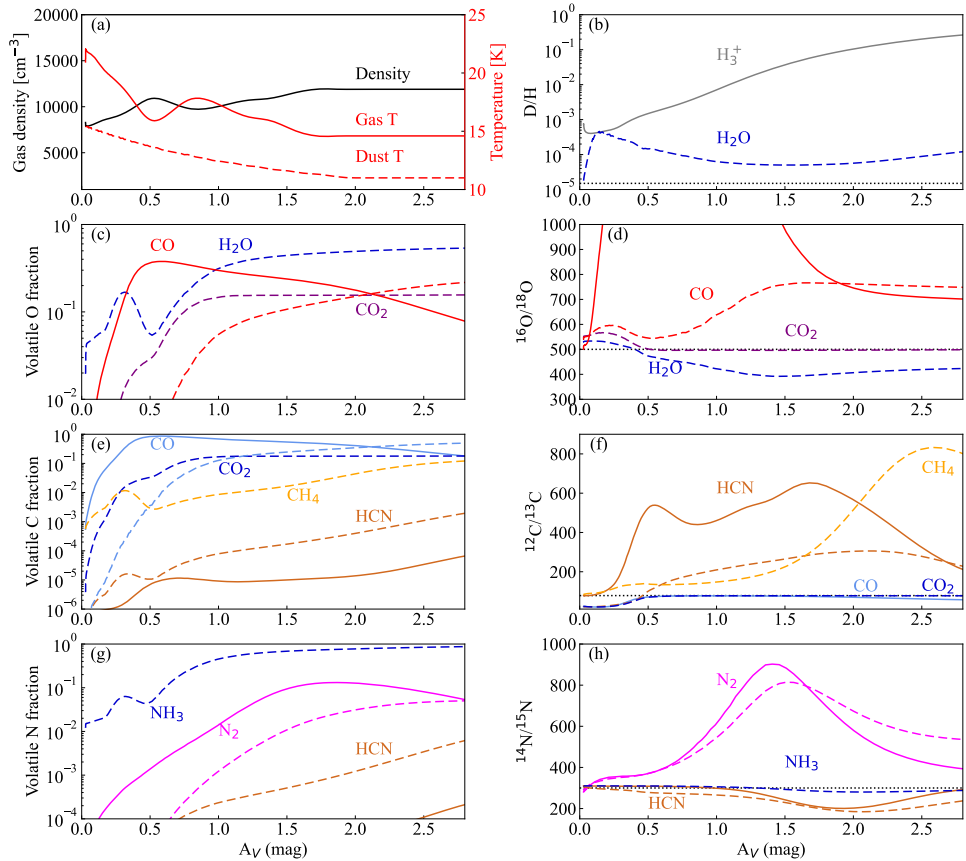}
\caption{Chemical evolution in a molecular cloud predicted by a gas-grain astrochemical model with H, C, N, O isotope fractionation chemistry 
{
\citep[][with some updates]{furuya18}.
}
Panel (a) shows physical evolution as a function of visual extinction ($A_{\rm V}$). 
Panels (c), (e), and (g) show the fraction of volatile elements locked up in selected chemical species, while
panels (b), (d), (f), and (h) show the degree of isotope fractionation.
Solid lines represent gas-phase species, while dashed lines represent icy species.
Black dotted lines in the right panels show the reference (elemental) isotope ratios adopted in the model.}
\label{fig:cloud_model} 
\end{figure*}

This section reviews the basic processes of isotope fractionation of
volatile elements (H, C, N, and O) in star- and planet forming regions.
As a guide, Figure \ref{fig:cloud_model} shows 
the chemical evolution in a forming and evolving molecular cloud predicted 
by a gas-grain astrochemical model with H, C, N, O isotope fractionation chemistry.
The physical model of the cloud is based on the 1-D shock model of \citet{bergin04}, which mimics the scenario that molecular clouds are formed due to the compression of diffuse HI gas 
by super-sonic accretion flows \citep{inutsuka15}.
As time proceeds, the column density of post-shock materials (i.e., molecular cloud) increases, which assists molecular formation by attenuating the interstellar UV radiation.
As such, the horizontal axes in Fig. \ref{fig:cloud_model} can be read as time ($A_v$ = 1 mag corresponds to $\sim$3 Myr in this particular model).
The gas density and temperature of the cloud is $\sim$10$^{4}$ cm$^{-3}$ and 10-20 K, respectively (Fig \ref{fig:cloud_model}a).
The chemical model is based on the gas-grain model presented by \citet{furuya18},
where H and N isotope fractionation were studied, and 
is extended to include O and C isotope fractionation chemistry.
In general, there are two mechanisms to cause isotope fractionation in gas-phase molecules:
isotope exchange reactions and isotope selective photodissociation.
Both mechanisms are included in the model, with up to date rate coefficients 
{
\citep[e.g.,][]{roueff15,loison19b,colzi20,loison20}.
As explained below, the dominant fractionation pathways of H and C isotopes are triggered by isotope exchange reactions (Sections \ref{sec:cloud_H} and \ref{sec:cloud_c}), while those for O and N isotopes are triggered by isotope selective photodissociation (Section \ref{sec:cloud_O} and \ref{sec:cloud_N}).
}


\subsection{Deuterium fractionation} \label{sec:cloud_H}
In star- and planet-forming regions, hydrogen and deuterium are primarily present as \ce{H2} and HD, respectively.
Deuterium fractionation can be understood as
the process that redistributes deuterium from HD into other species {
\citep[see also][for the review of deuterium fractionation chemistry]{ceccarelli14}.
}
In molecular clouds/cores, deuterium fractionation is mainly triggered by the isotope exchange reaction,
H$_3^+$ + HD $\rightleftarrows$ H$_2$D$^+$ + \ce{H2} \citep[e.g.,][]{watson76,dalgarno84}. 
The backward reaction is endothermic by 0-256 K {(the actual value differs depending on the nuclear spin state of \ce{H2}: ortho-\ce{H2} or para-\ce{H2}, see below), so thermal energy is required for it to proceed \citep{hugo09}.}
Thus the rate coefficient of the backward reaction is much lower 
than that of the forward reaction at low gas temperatures ($\lesssim$30 K).
This leads to H$_2$D$^+$ becoming more abundant over time with respect to H$_3^+$.
Since the H$_3^+$ isotopologues play a central role in the production of a variety of interstellar molecules \citep{herbst73}, 
the deuterium enrichment in H$_3^+$ is transferred to all the other molecules in the gas phase and on grain surfaces (Fig. \ref{fig:cloud_model}).{
For example, \ce{N2D+}/\ce{N2H+} ratio reaches to $>$0.1 in prestellar cores \citep[see][for the review of the observations of deuterated species towards prestellar cores]{ceccarelli14}.}
Chemical species that destroy H$_2$D$^+$ (e.g., CO and \ce{N2}) are readily frozen out onto cold dust grain surfaces, further enhancing the deuterium enrichment at low temperatures.
This is consistent with the higher D/H ratio in methanol (\ce{CH3OH}) than water observed in the warm gas ($>$100 K) around protostars, where ices have sublimated;
methanol is formed via a sequential hydrogenation of CO on grain surfaces, 
while water is predominantly formed via a sequential hydrogenation of O on grain surfaces prior to the CO freeze-out \citep{cazaux11,taquet14,furuya16}.

At warm temperatures ($\gtrsim$30 K), other exchange reactions, such as 
CH$_3^+$ + HD $\rightleftarrows$ \ce{CH2D+} + \ce{H2}, which have higher endothermicity ($\lesssim$500 K, again the value depends on the nuclear spin state of \ce{H2}) than the H$_2$D$^+$ pathway,
become the main driver of deuterium fractionation \citep{millar89,roueff13,nyman19}.
Spatially resolved observations of deuterated molecules in protoplanetary disks with ALMA have indicated that both the H$_2$D$^+$ pathway and the higher-temperature pathways are operating in disks 
\citep[e.g.,][ see Section \ref{sec:disk_D}]{huang17,salinas17,cataldi21}.

In addition to the gas temperature, the spin temperature of \ce{H2} (\emph{i.e.}, the ortho-to-para {abundance} ratio; OPR) is another key quantity that controls deuterium fractionation \citep{pagani92,flower06,taquet14,sipila17}.
The internal energy of ortho-\ce{H2} is higher than that of para-\ce{H2} by 170.5 K, 
which helps to overcome the endothermicity of the backward exchange reactions.
Even when the \ce{H2} OPR is small, $\sim$10$^{-3}$, \ce{H2} rather than CO is the main destroyer of  \ce{H2D+} and thus deuterium fractionation can be slowed down \citep{furuya15}.
When \ce{H2} is formed by grain surface reactions, the \ce{H2} OPR corresponds to the statistical value of three \citep{watanabe10}, but its value can fall over time via nucluear spin conversion in the gas phase, through proton transfer reactions (e.g., o-\ce{H2} + \ce{H+} $\rightleftarrows$ p-\ce{H2} + \ce{H+}) \citep{hugo09,honvault11}, and on grain surfaces \citep{ueta16,tsuge19,tsuge21}.
{Astrochemical models with nucluear spin conversion of \ce{H2} at the formation stage of molecular clouds predict that the \ce{H2} OPR is already much lower than the statistical value of three, when the main component of the gas becomes \ce{H2} molecules \citep{furuya15,lupi21}.}
\ce{H2} molecules are not easily observable, and 
{the \ce{H2} OPR has been estimated 
from observations of molecules other than \ce{H2}, such as \ce{N2D+} and \ce{H2D+}, as their abundances depend on the the \ce{H2} OPR.}
Observational studies have so far consistently suggested that the \ce{H2} OPR is much lower than unity (10$^{-4}$--0.1) in the cold gas of prestellar cores and protostellar envelopes \citep[e.g.,][]{pagani11,brunken14,harju17}.



Deuterium fractionation occurs primarily as a result of gas phase processes, 
but grain surface reactions at low temperatures ($\sim$10 K) can {further} enhance the level of fractionation \citep{watanabe08,fedoseev15}.
The degree of deuterium fractionation of formaldehyde
and methanol can be enhanced by the substitution and abstraction reactions of H and
D atoms at $\sim$10 K after their formation \citep{watanabe08,taquet12,cooper19}.
As other examples, methylamine (\ce{CH3NH2}) and ethanol (\ce{CH3CH2OH}) are also subject to the
H/D substitution reactions \citep{oba14,oba16}.
Moreover, the efficiency of the substitution
and abstraction reactions depends on the molecular functional group, e.g., deuterium substitution can be effective in the
\ce{CH3}--group of methanol but less so in its OH--group.
This can result in a high \ce{CH2DOH}/\ce{CH3OD} ratio ($>$1), as observed 
in the warm gas of embedded low-mass protostars \citep{taquet12}.
In contrast, once water is formed, it does not efficiently react with D atoms to be deuterated \citep{nagaoka05}.
{
In addition, several laboratory studies showed that H--D exchanges between hydrogen-bonded molecules in mixed ices occur efficiently at warm temperatures ($\gtrsim$70 K) \citep{ratajczak09,faure15,lamberts15}.
See the discussion in \citet{furuya16} for further details.
}


In the innermost regions of protoplanetary disks, where gas temperatures are greater than $\sim$500 K, 
isotope exchange reactions between deuterated species with \ce{H2}, 
such as HDO + \ce{H2} $\rightleftarrows$ H$_2$O + HD and \ce{CH3D} + \ce{H2} $\rightleftarrows$ \ce{CH4} + HD (with activation energy barriers of 5170 K and 4600 K, respectively), 
could reduce the degree of dueteration \citep{richet77,lecluse94}.
Presently, however, there are no observational constraints on the D/H ratio of water in protoplanetary disks.

\subsection{Oxygen isotope fractionation} \label{sec:cloud_O}
In contrast to deuterium fractionation, which is powered by isotope exchange reactions, 
oxygen isotope fractionation is thought to be caused by
isotope selective photodissociation of CO, followed by the formation of water ice \citep[e.g.,][]{clayton02,yurimoto04,lyons05}.
CO photodissociation is subject to self-shielding (i.e., CO itself can be the dominant absorber of dissociating UV radiation instead of dust grains).
Because rarer isotopologues (e.g., \ce{C^18O} and \ce{C^17O}) are less abundant, they are not self-shielded until deeper into the cloud or the disk.
This makes CO photodissociation an isotope-selective process \citep[e.g.,][]{visser09,miotello14}.{
See also \citet{chakraborty17} for the experiments of photodissociation of the CO isotopologues.}
The underabundance of \ce{C^18O} with respect to \ce{^12CO} or \ce{^13CO} due to the isotope selective photodissociation 
has been observed in molecular clouds illuminated by UV photons from nearby stars \citep[e.g.,][]{shimajiri14,chong18,yamagishi19} and in a protoplanetary disk \citep{smith09}.
As a result of isotope selective photodissociation of CO, $^{17, 18}$O-enriched atomic oxygen is produced 
and can be converted into water ice.
Then the two major reservoirs of volatile oxygen, CO and water can have different isotope compositions;
the former is depleted in heavy oxygen isotopes, while the latter shows the opposite trend (see Fig. \ref{fig:cloud_model}d).
There is no constraint on oxygen isotope fractionation of water in star- and planet-forming regions.
Cometary water, which could originate from the parent molecular cloud of the solar system \citep[e.g.,][]{cleeves14,altwegg17}, is enriched in \ce{^18O} (see Section \ref{sec:comet_O}).

Oxygen isotope compositions of other minor O-bearing species (i.e., neither CO nor \ce{H2O}) would depend on the source of oxygen from which they form.
For example, it is expected that \ce{CH3OH} and \ce{H2CO} are depleted in heavy oxygen isotopes, as they are formed via sequential reactions
of atomic hydrogen with CO on grain surfaces \citep[e.g.,][]{watanabe02,fuchs09}.
Note that in contrast to deuterium fractionation, there is no clear evidence that grain surface reactions modify the degree of oxygen isotope fractionation \citep{cooper19}.
CO$_2$, which is mainly formed via CO + OH $\rightarrow$ CO$_2$ + H on grain surfaces,
is expected to show no oxygen isotope fractionation, as OH is formed 
by the photodissociation of water ice and/or the hydrogenation of atomic oxygen (dashed orange line  in Fig. \ref{fig:cloud_model}).

\citet{loison19b} explored the impact of isotope exchange reactions on oxygen isotope fractionation under the physical conditions appropriate for dense molecular clouds.
They found that the gas-phase O-bearing species could be {fractionated} by isotope exchange reactions, 
such as \ce{^18O} + XO $\rightleftarrows$ O + \ce{X^18O} + $\Delta E$, where X is either O, N, or S 
and $\Delta E \approx  30-40$ K, 
when atomic oxygen is moderately abundant ($\gtrsim$10$^{-5}$ with respect to \ce{H2}).
The astrochemical model of \citet{loison19b} successfully reproduced the \ce{^18O} enrichment in SO observed 
in cold dense cores.
It should be noted, however, that high abundance of atomic oxygen in the gas phase leads to the formation of \ce{O2} \citep[e.g.,][]{hincelin11}, while the detection of \ce{O2} in cold clouds is rare \citep[][]{goldsmith00,pagani03,liseau12}.
A unique feature of oxygen is that it has three stable isotopes whereas hydrogen, carbon and nitrogen have only two stable isotopes. As a result, in principle, effects by isotope selective reactions could be resolved from those by isotope exchange reactions by comparing the \ce{^18O}/\ce{^16O} ratio with the \ce{^17O}/\ce{^16O} ratio (see Section \ref{sect:meteorites}).

In very inner regions of protoplanetary disks, where gas temperature is greater than $\sim$500 K,
oxygen isotope exchange can occur between silicates and \ce{H2O} gas (see Section 6).

\subsection{Carbon isotope fractionation} \label{sec:cloud_c}
Carbon isotope fractionation occurs via isotope selective photodissociation of CO \citep[e.g.,][]{visser09} and an exothermic isotope exchange reaction:
CO + \ce{^13C+} $\rightleftarrows$ \ce{^13CO} + \ce{C+} + 35 K \citep[e.g.,][]{watson76,langer84}.
PDR models have predicted that the former is less important than the latter, 
because the former is efficient only at $\sim$1 $A_{\rm V}$ \citep{rollig13}.
Astrochemical models under dark cloud physical conditions have predicted 
that carbon-bearing molecules are divided into two groups from the view point of \ce{^13C} fractionation via the isotope exchange reaction \citep[e.g.,][]{langer84}: 
CO and molecules formed from CO (e.g., \ce{CO2} and \ce{CH3OH}) are enriched in \ce{^13C}, while molecules formed from \ce{C+} (e.g., HCN, CN and \ce{CH4}) are \ce{^13C}-poor (Fig. \ref{fig:cloud_model}f). 
However, because CO is the main reservoir of volatile carbon, the \ce{^13C} enrichment of CO cannot be large.
On the other hand, \ce{HCO+} is affected by another exchange reaction, \ce{^13CO} + \ce{HCO+} $\rightleftarrows$ \ce{CO} + \ce{H^13CO+} + 17.8 K \citep{langer84,mladenovic17}, and consequently, \ce{HCO+} becomes further enriched in \ce{^13C} compared to CO.

The above mentioned scenario is challenged by the measurements of \ce{^12C}/\ce{^13C} ratios of CN and HCN in dense clouds,
which suggest these nitriles are enriched in \ce{^13C} \citep[][Fig. \ref{fig:ratios}]{daniel13,magalhaes18}.
Recent quantum chemical calculations have identified several other isotope exchange reactions, such as
\ce{^13C} + HCN $\rightleftarrows$ C + \ce{H^13CN} + 48 K and \ce{^13C} + C$_3$ $\rightleftarrows$ C + \ce{^13CC2} + 28 K \citep{roueff15,colzi20,loison20}.
The newly identified exchange reactions modify the classical picture mentioned above;
nitriles and carbon-chain molecules can be enriched in \ce{^13C} 
depending on physical environments as demonstrated by the gas-grain astrochemical model of dense cores with $A_{\rm V} = 10$ mag in \citep{loison20}.
Although these new isotope exchange reactions are taken into account in the model shown in Fig. \ref{fig:cloud_model},
HCN is depleted in \ce{^13C}, likely because the model only considers the regions where $A_{\rm V}$ is not high ($<$3 mag) and the interstellar radiation field is not fully shielded.

Oxygen-bearing complex organic molecules (COMs, e.g., glycolaldehyde and dimethyl either) in the warm gas around protostars IRAS 16293-2422, where ices have sublimated are 
enriched in \ce{^13C} by up to a factor of 2 compared to the elemental ratio in the local ISM \citep[][Fig. \ref{fig:ratios}]{jorgensen16,jorgensen18}.
COMs are thought to form on warm ($\gtrsim$20 K) dust grains irradiated by UV \citep[e.g.,][]{herbst09}.
The enrichment of \ce{^13C} in COMs might be due to slightly higher binding energy of \ce{^13CO} than \ce{^12CO} (i.e., resident time of \ce{^13CO} on grains is longer than \ce{^12CO}) and/or the effect of stellar UV radiation, which produces \ce{^13C} enriched atomic C by isotope selective photodissociation of CO \citep{jorgensen18}. 

In addition to the measurements of \ce{^12C}/\ce{^13C} ratio in various molecules, observations in molecular clouds have found that the abundances of the \ce{^13C} isotopomers 
{
(i.e., species with the same chemical formula, but differing in the position of \ce{^13C}) 
}
depend on which carbon atom in a molecule is substituted by \ce{^13C} 
\citep[e.g.,][]{takano98,sakai07,sakai10,taniguchi19,giesen20}.
For example, the abundance ratio of  \ce{^13CCH} to \ce{C^13CH} is greater than unity 
(i.e., more {stable} isotopomer \ce{^13CCH} is more abundant than \ce{C^13CH}) in dense molecular clouds \citep{sakai10,taniguchi19}.
These isotopomer ratios provide a clue of the formation pathway of the molecules or indicate the efficient position exchange reaction, which converts less stable isotopomers to more stable ones at low temperatures \citep[e.g., see][for the case of CCH]{sakai10,furuya11}.

\subsection{Nitrogen isotope fractionation} \label{sec:cloud_N}
\label{sec:n_frac}
According to observations towards dense clouds, there is no clear evidence of nitrogen isotope fractionation in the gas phase nitriles \citep[][and references therein]{hilyblant20},
while ammonia and \ce{N2H+} in the gas phase show 
significant \ce{^15N} depletion by a factor of $\sim$2--3 compared to the elemental ratio in the local ISM \citep[][Fig. \ref{fig:ratios}]{bizzocchi13,redaelli18}.
The significant depletion of \ce{^15N} in \ce{N2H+} indicates that its parent molecule, \ce{N2},
is also {depleted in \ce{^15N}}.
To the best of our knowledge, there are no direct measurements of the \ce{^14N}/\ce{^15N} ratio in any molecules in the hot ($>$100 K) gas, where ices have sublimated.
Therefore, the \ce{^14N}/\ce{^15N} ratio of molecular ices in the ISM is presently not constrained.
On the other hand, observations of comets in our solar system have found that HCN and \ce{NH3} in cometary ices are enriched in $^{15}$N compared to the solar wind, by a factor of $\sim$2--3 (see Section 4.3).

Compared to the other volatile elements, the mechanism of nitrogen isotope 
fractionation remains an open question.
Previously, it had been suggested that nitrogen isotope exchange reactions are efficient at low temperatures ($\sim$10 K), so gas-phase nitriles and ammonia can be enriched in $^{15}$N, and the gas-phase enrichment transferred to the ice \citep[$\sim$10 K; e.g,][]{terzieva00,charnley02,wirstrom12}.
However, recent reinvestigations of the exchange reactions by quantum chemistry calculations 
and astrochemical simulations have found that  fractionation by exchange reactions 
is much less efficient than had previously {thought} even at 10 K \citep{roueff15,wirstrom18,loison19a}.
The updated models are consistent with observations in dense clouds that 
the degree of nitrogen isotope fractionation is generally small.
However, a major puzzle remains regarding the depletion of \ce{^15N} in \ce{N2H+}; this is difficult to explain using isotope-exchange reactions, which should produce the opposite trend.
To explain the \ce{^15N} depletion of \ce{N2H+}, \citet{loison19a,hilyblant20} have proposed the possibility that the rates of dissociative recombination (\ce{N2H+} + $e^-$ $\rightarrow$ \ce{N2} + H) for \ce{N^15NH+} and \ce{^15NNH+} are larger than that for \ce{^14N2H+} (see the original papers for details).
This reaction is one of the main destruction pathways of \ce{N2H+}.
As noted by those authors, there are no experimental measurements of the dissociative recombination rates for \ce{N^15NH+} and \ce{^15NNH+} at low temperatures ($\sim$10 K). Experimental or/and  theoretical studies will therefore be required to verify the scenario. 
In the model in Fig. \ref{fig:cloud_model}, the two relevant rate coefficients are assumed to be the same.

Alternatively, nitrogen isotope fractionation could be induced by isotope selective photodissociation of N$_2$,
followed by the formation of \ce{^15N} enriched N-bearing icy species, 
such as \ce{NH3} and HCN ices \citep[][Fig. \ref{fig:cloud_model}h]{furuya18}.
{
See \citet{heays14} and \citet{chakraborty14,chakraborty16} for theory and experiments of the isotope selctive photodissociation of \ce{N2}, respectively.
}
This mechanism is an analog of oxygen isotope fractionation induced by isotope selective photodissociation of CO, followed by \ce{H2O} ice formation (compare Fig. \ref{fig:cloud_model}d and \ref{fig:cloud_model}h).
A model by \citet[][]{furuya18}, which simulates the nitrogen isotope fractionation during the formation and evolution of a molecular cloud via the compression of diffuse atomic gas,
could qualitatively explain the \ce{^15N} depletion in \ce{N2H+} and the \ce{^15N} enrichment in cometary ices (assuming cometary volatile was inherited from the ISM), but could not reproduce them quantitatively; 
the model predicted that the degree of fracitonation is up to several tens of per-cent.
However, their model is based on a specific physical situation.
More numerical studies, and varying physical parameters, are necessary to 
better understand the \ce{^15N} observations.
We note that if the isotope selective photodissociation is the main cause of nitrogen isotope fractionation, 
\ce{N2} would need to be the main gas-phase reservoir of elemental nitrogen; otherwise isotope selective photodissociation of \ce{N2} does not work \citep{furuya18}.
Then, the uncertainty in the mechanism of the nitrogen isotope fractionation is linked to our uncertainty regarding the main nitrogen reservoirs in the ISM.



\section{\textbf{PROTOPLANETARY DISKS}}
\label{sect:disks}

\begin{figure*}[t]
\centering
\includegraphics[width=1.0\linewidth]{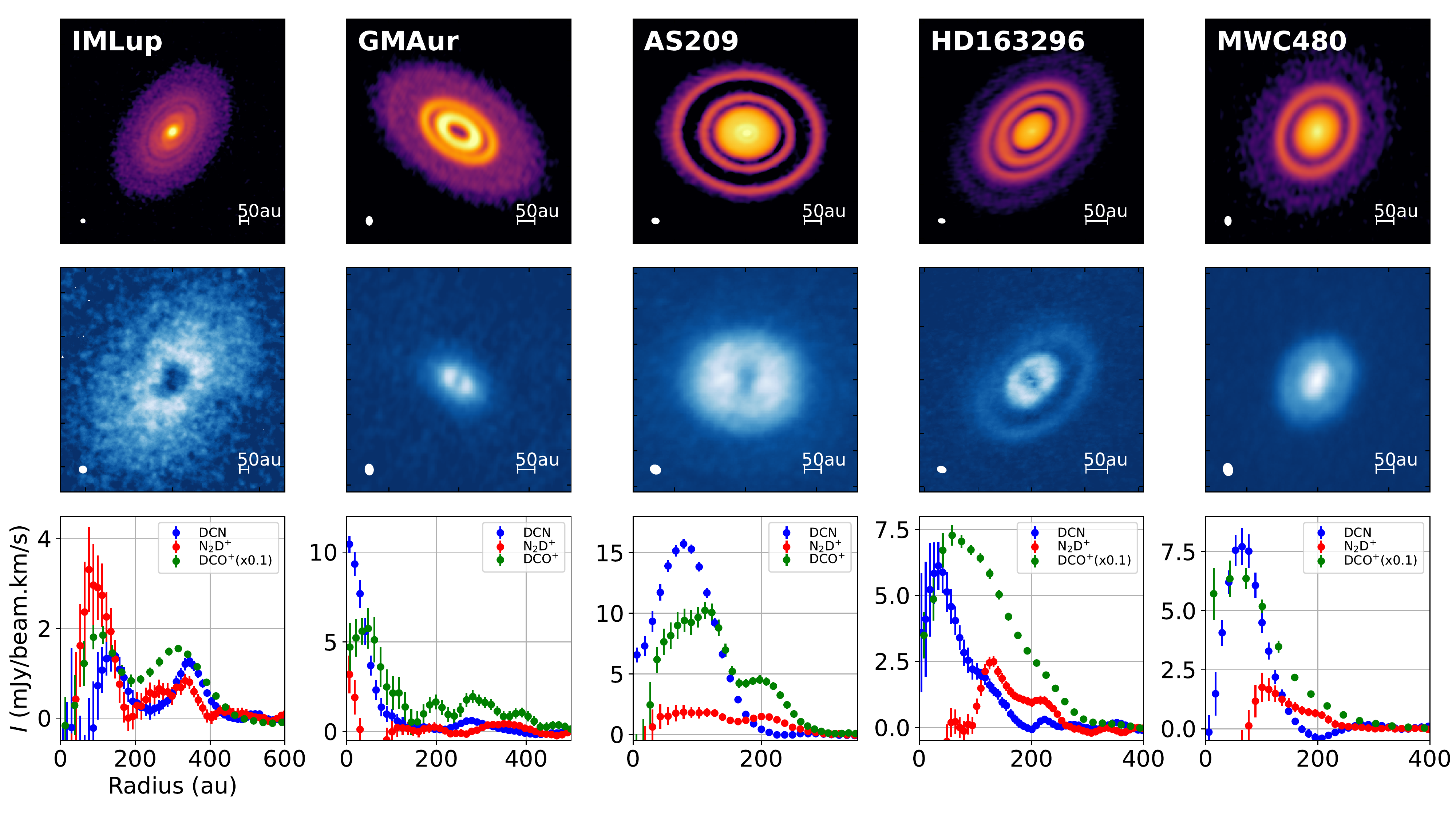}
\caption{Dust continuum images (top), integrated line images of HCN 3-2 (middle), radial profiles of deuterated species (bottom) towards five bright protoplanetary disks \citep[][]{oberg21c, cataldi21}. DCN, DCO$^+$, and \ce{N_2D^+} are located from inner to outer regions in this order for most disks.}
\end{figure*}

Chemistry in protoplanetary disks can be described by analogy with that of molecular clouds. Isotope fractionation due to isotope exchange reactions proceeds in the cold outer regions of disks, while isotope selective photodissociation takes place in the surface layer of the disks, which is directly irradiated by stellar UV radiation. The dust grains in disks are the building {blocks} of planets/asteroids/comets in planetary systems. They collisionally grow and move in vertical and radial directions inside the disks; settling towards the disk midplane and radially migrating towards/away from the central stars, depending the direction of gas pressure gradient force. This differs from the situation in molecular clouds where dust grains are small and coupled with the gas motion. Isotope ratios could trace how and where the objects in planetary systems are formed in disks.

\subsection{Cold and warm paths of deuteration in disks} \label{sec:disk_D}

HCO$^+$ and HCN have the strongest emission lines of hydrogen containing species towards protoplanetary disks. The deuterated species, DCO$^+$, was initially detected using single dish telescopes \citep{vandishoeck03,guilloteau06}, and a tentative detection of DCN by JCMT was confirmed by Submillimeter Array (SMA) observations \citep{thi04,qi08}. The subsequent SMA and IRAM PdBI survey observations detected DCO$^+$ towards more objects \citep[e.g.,][]{oberg10,oberg11,teague15} and now ALMA has revealed detailed spatial distributions of both DCO$^+$ and DCN in disks.

Since deuterium fractionation is driven by isotope exchange reactions (see Section 2.1), the enhancement of the D fractionation in the cold outer disks has been theoretically suggested and observationally confirmed especially for DCO$^+$ \citep[e.g.,][]{aikawa99,aikawa01,qi08}. On the other hand, HCO$^+$ and DCO$^+$ deplete if gas-phase CO is frozen out on grains since HCO$^+$ is formed by protonation of CO. Therefore, it is expected that DCO$^+$ is enhanced most around the CO snowline inside which CO exists in the gas-phase without freezing on dust grains, and ALMA observations with high spatial resolution have actually revealed it \citep[e.g.,][]{mathews13,salinas17}. In addition, ring-like enhancements of DCO$^+$ in the outer disk are explained by photodesorption of CO \citep[e.g.,][]{oberg15}. Meanwhile, ALMA observations also show that both DCN and DCO$^+$ exist in the inner disk, with DCN distributed in more inner disks compared with DCO$^+$, suggesting warm formation paths for deuterated species \citep[e.g.,][, see Figure 3]{huang17,kastner18,oberg21b,cataldi21}. Theoretical models show that DCN and DCO$^+$ can be formed through, for example, \ce{N + CHD/DCO} $\rightarrow$ \ce{DCN + H/O} and \ce{CH_4D^+ + CO} $\rightarrow$ \ce{DCO^+ + CH_4}, respectively  in addition to the cold path induced by \ce{H_3^+ + HD} $\rightleftarrows$ \ce{H_2D^+ + H_2} \citep[e.g.,][]{cleeves16,aikawa18}.

\ce{N_2D^+} has also been  detected in protoplanetary disks by ALMA \citep{huang17,salinas17,cataldi21}. \ce{N_2D^+} is mainly distributed in the outer disks (Figure~3), consistent with the theoretical models which predict that only the cold pathway is efficient for  \ce{N_2D^+} formation. High deuterium fractionation of \ce{N_2D^+}$/$\ce{N_2H^+}$\sim 0.1-1$, compared with \ce{DCO^+}$/$\ce{HCO^+} and DCN$/$HCN$\sim 0.01-0.1$, also reflects that \ce{N_2D^+} is formed only in cold outer disks. \ce{C_2D} is another deuterated species recently detected in protoplanetary disks by ALMA \citep{loomis20}. \ce{C_2D} possibly has a radial distribution similar to that of \ce{DCO^+}. 

HD has been detected towards protoplanetary disks by Herschel Space Observatory \citep{bergin13,mcclure16}. Since \ce{H2} is the dominant reservoir of hydrogen, HD is thought not to be fractionated, maintaining its cosmic abundance of $\sim 10^{-5}$. As \ce{H2} line emission mainly traces relatively hot surface regions of disks, the HD lines provide a useful estimate for disk gas masses \citep[e.g,][]{trapman17,kama20}.

Since \ce{H_2D^+} is a key species which initiates the deuterium chemistry in the cold outer disks, there have been attempts to observe the species towards protoplanetary disks. However, observations failed to detect both \ce{H_2D^+} and \ce{D_2H^+} \citep[e.g.,][]{qi08,chapillon11}, possibly due to the concentration of D in \ce{D_3^+} as is suggested by theoretical models \citep[e.g.,][]{willacy09}. HDO is another important deuterated species which has not been detected in protoplanetary disks so far \citep[e.g.,][]{guilloteau06}, while it has been studied very well by theoretical models \citep[e.g.,][]{furuya13,albertsson14,cleeves14}. Detection of deuterated water in disks is one of the important keys to understanding the origin of water in the solar system objects linking from molecular clouds.

\subsection{Nitrogen fractionation in disks}

\begin{figure}[t]
\centering
\includegraphics[width=1.0\linewidth]{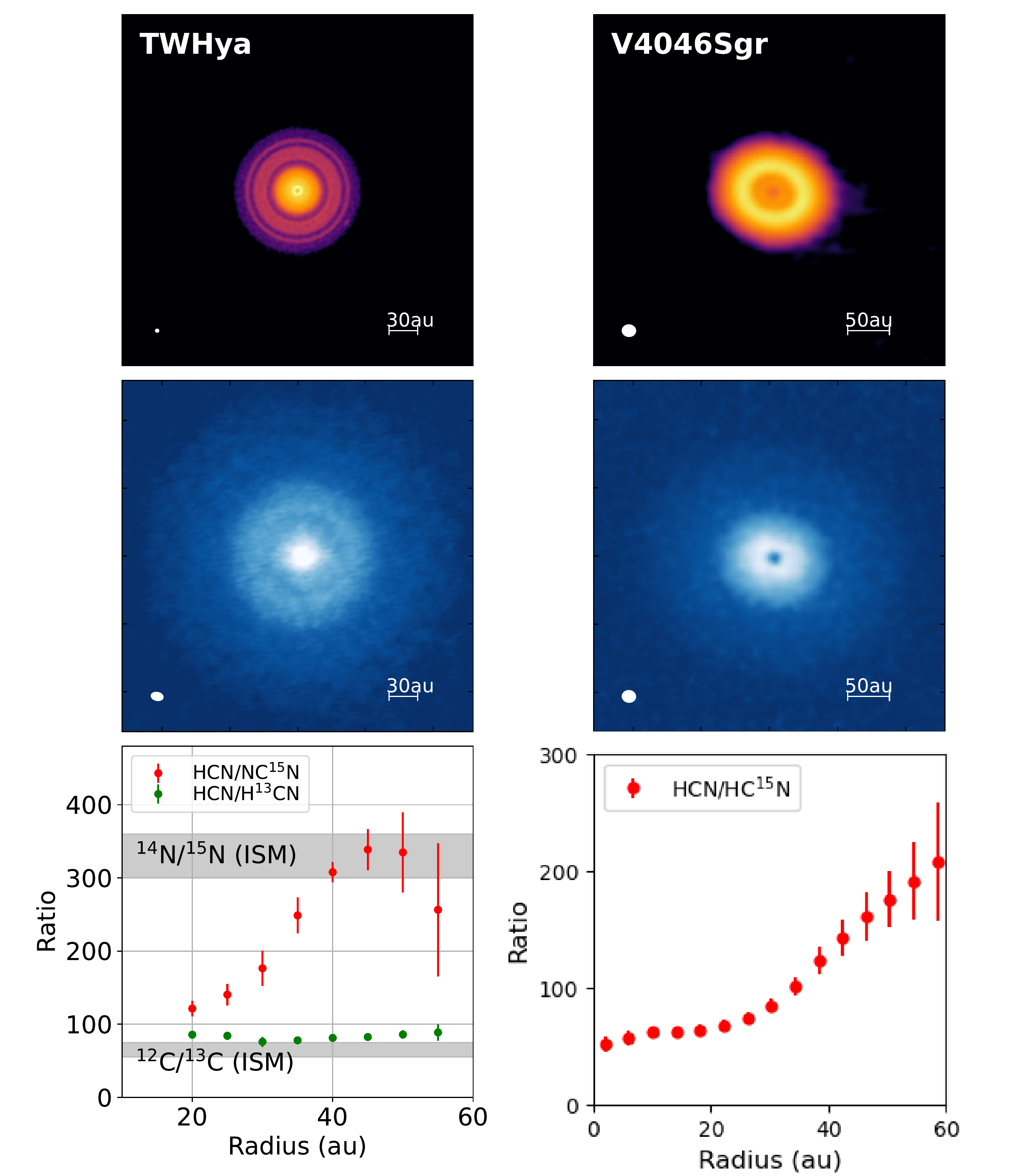}
\caption{Dust continuum images (top), total integrated intensity maps of HCN(4--3) (middle) and radial profiles of \ce{HC^{14}N}$/$\ce{HC^{15}N} (TW Hya, V4046 Sgr) and \ce{H^{12}CN}/\ce{H^{13}CN} (TW Hya) ratios (bottom) towards nearby protoplanetary disks \citep[cf.][]{tsukagoshi19,hilyblant19,guzman21}. The {\ce{HC^{14}N}$/$\ce{HC^{15}N} ratio} decreases in the inner regions for both disks.}
\end{figure}

Nitrogen isotopes were detected in protoplanetary disks for the first time by ALMA \citep{guzman15,guzman17,hilyblant17,hilyblant19,booth19}. The measured nitrogen isotope ratios, \ce{HC^{14}N}$/$\ce{HC^{15}N}, are similar to those in protostellar envelopes and comets, and \ce{C^{14}N}$/$\ce{C^{15}N} has a slightly higher value (see Figure~1). Spatially resolved \ce{HC^{14}N}$/$\ce{HC^{15}N} measurements have been obtained towards two nearby objects: TW Hya and V4046 Sgr, which show that the ratio decreases inside $\sim 50$ au for both objects (Figure~4). Meanwhile, \ce{HC^{14}N}$/$\ce{HC^{15}N} has been measured towards disks with various ages from Class~0/I to Class~II, suggesting that the ratio decreases with evolutionary stage. This could reflect an increased isotope selective photodissociation efficiency as the surrounding envelope clears \citep{bergner20}.

Nitrogen fractionation models in protoplanetary disks have been constructed \citep{visser18,lee21}. As is shown in Section 2, nitrogen fractionation is mainly caused by isotope selective photodissociation rather than isotope exchange reactions. In the case of protoplanetary disks, the fractionation therefore proceeds mainly in the surface layer and the outer edge of the disks. Observations suggest that no {clear} correlation exists between the degree of deuterium and nitrogen fractionation, which is consistent with the model that deuterium and nitrogen fractionation are mainly caused by different mechanisms (exchange reactions vs. isotope selective photodissociation) \citep{guzman17}. 
The fractionation depends on the properties of dust grains, that is, how much ultraviolet irradiation can penetrate deeper in the disks. {The fractionation becomes significant if small dust grains are more abundant and the molecular layer shrinks \citep{lee21}.} The observed tendency for \ce{HC^{14}N}$/$\ce{HC^{15}N} to increase in the outer parts of disks could then be explained if the small dust grains deplete in the outer disk, as suggested by dust evolution models \citep[e.g.,][]{birnstiel12}.

\subsection{Carbon and oxygen fractionation in disks}

Isotopologues of CO (\ce{^{12}C^{16}O, ^{13}C^{16}O, ^{12}C^{18}O}), were the first molecules detected in protoplanetary disks, using millimeter arrays and a single dish telescope \citep{kawabe93,koerner93,guilloteau94}. They emit some of the strongest lines in disks, and these lines are commonly used to survey the gaseous component of disks \citep[e.g.,][]{williams14,ansdell16}. However, since \ce{^{12}C^{16}O} lines are optically thick, it is difficult to measure the isotope ratios, using these lines. Together with observations of \ce{^{13}C^{16}O, ^{12}C^{18}O, ^{12}C^{17}O} lines and multi-transition lines of \ce{^{12}C^{16}O}, CO isotope ratios 
are measured toward the disk around HD~163296 by fitting observations to model calculations \citep{qi11}. The derived values are consistent with the ISM values. Another measurement has been taken toward a protoplanetary disk around VV~CrA, using high-dispersion infrared spectroscopic observations of CO ro-vibrational line absorption \citep{smith09}. The measured values are \ce{^{12}C^{16}O}$/$\ce{^{12}C^{18}O} $=690\pm 30$,  \ce{^{12}C^{16}O}$/$\ce{^{12}C^{17}O} $=2800\pm 300$, and
\ce{^{12}C^{18}O}$/$\ce{^{12}C^{17}O} $= 4.1\pm 0.4$, which are not significantly different from the ISM values.
ALMA and IRAM NOEMA have also made it possible to detect the rarer isotopologues \ce{^{13}C^{18}O} and \ce{^{13}C^{17}O} from protoplanetary disks \citep{zhang17,zhang20,booth19,booth20}. \ce{^{12}C}/\ce{^{13}C}$=40^{+9}_{-6}$ is suggested toward the TW~Hya disk by fitting model calculations to the observations. These observations of optically thin CO isotopologues are often used as mass tracers by simply assuming that there is no significant fractionation in carbon and oxygen isotopes. 

\ce{H^{13}CO^+} has been detected by single-dish and SMA observations toward a protoplanetary disks \citep{thi04,qi08}. ALMA has made it possible to spatially resolve both  \ce{H^{13}CO^+} and \ce{H^{13}CN} towards some disks \citep{guzman15,guzman17,huang17,booth19,hilyblant19}. Since the corresponding \ce{HCO^+} and HCN lines are optically thick however, it is fundamentally difficult to measure the isotope ratios directly. However, the disks become optically thinner at large radii, and a low \ce{H^{12}CO^+}$/$\ce{H^{13}CO^+} ratio is measured in the outer edge of the HD~97048 disk by fitting observations with model calculations \citep{booth19}. Also, taking advantage of optically thin hyper-fine lines of HCN, \ce{^{12}C}$/$\ce{^{13}C} in HCN is measured towards the TW Hya disk, showing \ce{^{12}C}$/$\ce{^{13}C}$ =86\pm 4$ on average, which is slightly higher than the ISM value, and the radial dependence is not high (from $\sim 75.9\pm 7.0$ to $\sim 88.7\pm 11.1$) throughout the disk inside 55 au --- significantly different from the \ce{HC^{14}N}$/$\ce{HC^{15}N} ratio \citep[][Figure 4]{hilyblant19}.
Carbon isotope models of protoplanetary disks show that carbon fractionation proceeds in the surface layer and outer disk where isotope selective photodissociation occurs \citep{woods09,visser18}.

{\ce{HC^{18}O+} has recently been detected towards the TW Hya disk by ALMA, and a ratio of \ce{H^{13}CO+}/\ce{HC^{18}O+}$ =8.3\pm 2.6$ was derived. Considering exchange reactions and the derived \ce{H^{13}CO+}/\ce{HC^{18}O+} ratio, \ce{^{13}CO}/\ce{C^{18}O}$ =8.1\pm 0.8$ is obtained, which is consistent with the detailed model calculation \citep{furuya22}.}
Oxygen isotope fractionation in the protosolar nebula has been proposed in the context of a possible link to the oxygen isotope anomaly discovered in meteorites \citep[][see also Section 6]{lyons05}. Relatively simple protoplanetary disk chemical models with isotope selective photodissociation of CO and isotope exchange reactions have been constructed, mainly focusing on fractionation of the CO isotopologues in order to help derive disk gas masses observationally \citep{miotello14,miotello16}. Further investigation of carbon and oxygen isotope fractionation in various molecules in disks will help provide insight into the possible link between materials in planet forming regions and those in the solar system.

\subsection{Future prospects to link planet forming regions to the solar system}

{In order to link planet forming regions to the solar system, we need to understand how observable gas-phase compositions in the disk surface are linked to the ice composition near the disk midplane (\emph{i.e.} the building blocks of planetesimals and then objects in planetary systems). Theoretically, constructing models of gas-phase and grain surface isotope chemistry together with turbulent motion and dust dynamics \citep[e.g.,][]{krijt20} will improve our understanding. Observations of isotopologues which are just evaporated from ice and preserve the ice isotopic ratios, for example, in the disks of FU Ori-type young outbursting objects \citep[e.g.,][]{lee19}, disks with strong dust concentrations \citep[e.g.,][]{vandermarel21,booth21}, and potentially in debris disks \citep[e.g.,][]{kospal13,moor17,higuchi19}, will also be key. 
Searching for potential correlations among different isotopes \citep[e.g.,][]{guzman17,marty12,alexander12} will also be useful for further understanding the link.}





\section{\textbf{COMETS}}
\label{sect:comets}

Comets are composed of ice and dust accreted during the epoch of planet formation, and are believed to have been subject to relatively little chemical alteration since that time \citep{mum11}. Their study therefore provides a unique window into the isotopic composition of material from the early solar system, including volatiles (ice) from the protosolar disk and prior interstellar cloud, as well as refractory material (dust) that may have undergone significant thermal processing closer to the sun. {The volatile inventory of comets is dominated by H$_2$O, CO$_2$ and CO ices, complemented by similar abundances of organic molecules and sulphur-bearing species to those found in interstellar ices, as well as in the warm, protostellar gas where ice mantles have been evaporated.}

\subsection{Carbon isotopes in the pre-Rosetta era}

Isotopic studies of comets, date back to the first measurement of the $^{12}$C/$^{13}$C ratio in C$_2$, observed in optical emission from the coma of Ikeya 1963 I \citep{sta64}. Observations of $^{13}$CC and C$_2$ in a number of subsequent apparitions revealed an average $^{12}$C/$^{13}$C ratio of $90\pm8$ (error weighted mean {with associated one-sigma uncertainty}; see review by \citealt{boc15}), which is fully consistent (within errors) with the Solar and terrestrial carbon isotopic abundances. Cometary CN was found to suffer from less spectral line blending than C$_2$, and a mean value of $^{12}$C/$^{13}$C = $91\pm4$ was derived from CN in a diverse sample of 21 comets by \citet{man09}.

\subsection{Deuteration of cometary volatiles}

The power of isotopic measurements for elucidating cometary origins became more apparent with the D/H measurement in comet 1P/Halley's water coma by the Giotto spacecraft ($(3.1\pm0.3)\times10^{-4}$; \citep{ebe95}, later revised to $(2.1\pm0.3)\times10^{-4}$; \citealt{bro12}). These mass spectrometry measurements indicated a significant deuterium enrichment in cometary ice compared with the primordial (Solar) value of $2.1\times10^{-5}$. Detections of D/H have now been published for 11 different comets using in situ mass spectrometry, ground-based optical spectroscopy of OD/OH and mm/sub-mm/IR measurements of HDO/H$_2$O/H$_2^{18}$O, revealing a distribution of values [D/H]$_{\rm H_2O}$ in the range (1.0--8.1)$\times10^{-4}$, with an error-weighted mean of $(2.1\pm0.1)\times10^{-4}$. This includes the value of $(5.3\pm0.7)\times10^{-4}$ measured by the ROSINA mass spectrometer in comet 67P \citep{alt15}, as well as the recently-obtained value $(1.6\pm0.7)\times10^{-4}$ in comet 46P using the Stratospheric Observatory for Infrared Astronomy (SOFIA; \citealt{lis19}). A summary of the cometary D/H measurements obtained to-date is shown in Figure \ref{fig:ratios}.

While there appears to be genuine diversity in [D/H]$_{\rm H_2O}$ across the comet population, the currently available statistics are relatively noisy, and constitute a sensitivity-limited sample. Some upper limits or non-detections are likely to have been unreported, and upper limits are not included in our average value, so the error-weighted mean is therefore biased in the direction of the (larger) reported D/H measurements. More accurate statistics from future observations will be very useful to confirm the true distribution (and average D/H ratio) in comets. For the time being however, it is reasonable to consider typical comets as somewhat enriched in deuterium relative to VSMOW --- 67P being the most significantly enriched, at (2.9--3.8) $\times$ VSMOW.

There is no strong evidence that different D/H values occur in comets from different dynamical reservoirs (originating from different parts of the protoplanetary disk). However, a trend for decreasing D/H as a function of active surface fraction (the ratio of sublimating area to the total surface area of the nucleus) observed by \citet{lis19} could be interpreted as resulting from the presence of two isotopically distinct ice reservoirs in the nucleus: one reservoir in the relatively ice-poor material residing close to the surface of the nucleus (preferentially released in lower-activity comets), and a different, lower D/H reservoir in the ice-rich grains, which can contribute strongly to the active (sublimating) surface area (see also \citealt{ful21}). Alternative explanations for the trend observed by \citet{lis19} include the possibility of sublimation-dependent deuterium fractionation, for example, as observed in the laboratory by \citet{lec17} (albeit at a relatively low level), or as a result of non-steady-state deuterium enrichment of the coma, as predicted by the sublimation model of \citet{pod02}.

The detection of a surprisingly large D$_2$O abundance in comet 67P by Rosetta (with D$_2$O/HDO = $(1.8\pm0.9)\times10^{-2}$) provided crucial new contraints on the origins of cometary H$_2$O, beyond what can be inferred from the HDO/H$_2$O ratio alone \citep{alt19}. The \citet{fur17} model for a collapsing, infalling, irradiated protostellar envelope generates insufficient deuteration to reproduce the 67P D$_2$O/HDO ratio. However, similar values of D$_2$O and HDO enrichment to those found in 67P readily occur in the cold interstellar phase preceding star formation. The doubly-deuterated isotopologue thus provides a strong indicator for the interstellar heritage of cometary water. 

Another important result from the Rosetta mission was the first detection of deuterated CH$_3$OH in comets, with a relatively large abundance ratio (CH$_2$DOH + CH$_3$OD)/CH$_3$OH = $(5.5\pm0.5)\times10^{-2}$ measured in the coma of 67P \citep{dro21}. The similarity between this value and the D$_2$O/HDO ratio, combined with the theoretical result that deuteration of both H$_2$O and CH$_3$OH proceeds in interstellar ices \emph{via} D-H atom-exchange reactions, implies that deuterated methanol formed relatively late in the protosolar cloud core (prior to collapse), when gas-phase atomic D was more abundant, but while the temperatures were still low enough to preserve significant amounts of CO in the ice for CH$_3$OH production. The relatively lower ratios of HDS/H$_2$S and NH$_2$D/NH$_3$ in comet 67P of $\sim6\times10^{-4}$ and $\sim1\times10^{-3}$, respectively \citep{alt19}, and of DCN/HCN ($2.3\times10^{-3}$) in comet O1 (Hale-Bopp) are consistent with deuteration ocurring in these molecules at a somewhat earlier (prestellar) epoch \citep{dro21}, when atomic D was less abundant.

\subsection{Nitrogen fractionation in comets}

Observations of nitrogen fractionation in comets were reviewed by \citet{boc15} and \citet{hilyblant17}, who reported a weighted average $^{14}$N/$^{15}$N ratio of $144\pm3$ in HCN, CN and NH$_2$ (from 31 comets).  Cometary ices observed to-date show a ubiquitous enrichment in $^{15}$N with respect to the bulk of the material from which the solar system formed (represented by the Jovian and solar wind-derived values: $^{14}$N/$^{15}$N $\approx440$; {\citealt{owen01,marty11}}), as well as with respect to the terrestrial value of 272 {\citep{junk58}}. There is no strong evidence for diversity in individual $^{14}$N/$^{15}$N values across the comet population (disk $^{14}$N/$^{15}$N ratios are also quite consistent, with values $\sim100$--300), implying a uniformity in the nitrogen fractionation processes occurring in the outer parts of protoplanetary disks, where comets are believed to have accreted. ROSINA mass spectromety of NH$_3$ and NO in the 67P coma provided in situ confirmation of this trend \citep{alt19}. Surprisingly, a consistent isotopic fingerprint was also found in 67P's molecular nitrogen, which had $^{14}$N/$^{15}$N $\sim130\pm30$. Considering N$_2$ is believed to be a major reservoir of nitrogen in protoplanetary disks \citep[e.g.,][]{walsh15}, and the giant planets (representative of that reservoir; see Section \ref{sec:planets}) show a lack of $^{15}$N fractionation, the molecular nitrogen observed in comet 67P is unlikely to have been directly inherited from primordial (or interstellar) N$_2$ in the disk, and must originate from more isotopically-processed nitrogenous material. Overall, the nitrogen isotopes in 67P are consistent with accretion from a relatively uniformly-fractionated ice reservoir, most likely dominated by the $^{15}$N-enrichment resulting from isotope-selective photodissociation of N$_2$ (see Section \ref{sec:n_frac}).

\subsection{Oxygen fractionation in cometary volatiles} \label{sec:comet_O}

\begin{figure}[h!]
\centering
\includegraphics[width=0.95\linewidth]{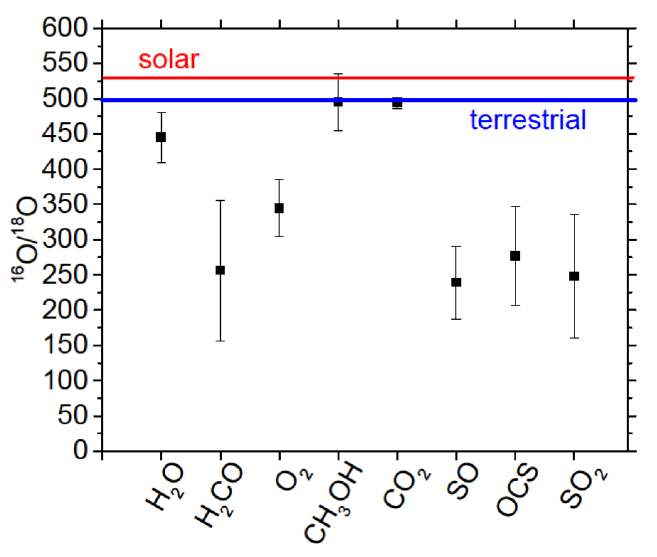}
\caption{$^{16}$O/$^{18}$O ratios for eight molecules observed using Rosetta mass spectrometry of the comet 67P coma. Solar and terrestrial baseline values are shown in red and blue, respectively. Figure reproduced from \citet{alt20}. \label{fig:o_comets}}
\end{figure}

Prior to the Rosetta mission to comet 67P, evidence for significant oxygen fractionation in comets was scarce, particularly when considering the weighted-mean $^{16}$O/$^{18}$O ratio of $495\pm19$ {(at $1\sigma$ confidence)} from the 9 measurements compiled by \citet{boc15}, which is consistent with the terrestrial value of 499 {and close to the solar value of 530; \citealt{RN4400}}. Rosetta measured  H$_2$O/H$_{2}${$^{18}$O} = $445\pm35$ in 67P \citep{sch19}, corresponding to a modest enhancement in $^{18}$O compared with terrestrial and solar wind values. Comet 67P's CO$_2$ (the next most abundant ice in the nucleus) on the other hand, shows no evidence for oxygen fractionation (with $^{16}$O/$^{18}$O = $494\pm8$). These oxygen isotope results are consistent with chemical models incorporating isotope-selective photodissociation of CO in the protosolar nebula. An $^{18}$O enrichment of 5-20\% is predicted in H$_2$O by the models of \citet{lyons05} and \citet{lee08} as a result of preferential photodissociation of C$^{18}$O compared with CO.  On the other hand, CO$_2$ forms from O-atom addition to the dominant CO isotopologue, which is less efficiently dissociated than the minor isotopologues due to self-shielding in the dense nebular gas.

The high sensitivity and resolution of Rosetta's ROSINA mass spectrometer afforded unprecedented precision in the $^{16}$O/$^{18}$O ratios for several previously unmeasured molecules in comets \citep{alt20}. O$_2$, H$_2$CO, and the sulphur-bearing species SO, SO$_2$ and OCS all show strong ($\sim$ factor of two) enrichments in $^{18}$O compared with the terrestrial value (see Figure \ref{fig:o_comets}), whereas CH$_3$OH shows a ratio consistent with terrestrial. Similarity of the $^{16}$O/$^{18}$O ratios in CO$_2$ and CH$_3$OH is expected based on their theorized common origin from atom-addition to CO on grain surfaces. By contrast, the discrepancy in $^{16}$O/$^{18}$O for H$_2$CO implies a different origin for this molecule; \emph{i.e.} CO cannot be its main chemical precursor. This is also implied by the strong ($\sim$ factor of two) enrichment in the $^{13}$C fraction of H$_2$CO, which is not observed in CO. CO$_2$, CH$_3$OH and C$_2$H$_6$ also show a lack of any significant $^{13}$C enrichment \citep{has17,alt20}. Observations of CO, H$_2$CO and CH$_3$OH isotopologues by  in the DR21 (OH) high-mass protostellar core show a similar trend to 67P, with H$_2$CO significantly enriched in $^{13}$C compared with CO and CH$_3$OH. Based on its unusual spatial distribution, cometary H$_2$CO is believed to originate (at least in part) from the degradation of organic-rich dust or macromolecular material in the coma \citep{cot08,cor14}, so the unusual isotopic composition of H$_2$CO in 67P could be inherited from organic refractories. Future observations of H$_2$CO isotopologues in comets with confirmed extended sources of H$_2$CO will enable further exploration of this hypothesis. 

The comet 67P results showing significant $^{18}$O enrichment in O$_2$, SO, SO$_2$ and OCS, along with modest enrichment in H$_2$O and negligible enrichment in CH$_3$OH can be explained as a result of isotopic exchange reactions involving atomic $^{18}$O in low temperature, interstellar gas \citep{loison19b} (also see Section \ref{sec:cloud_O}\rev{)}. Similar $^{18}$O-exchange reactions are also expected to occur in the cold gas in the outer regions of the protosolar disk {(Section \ref{sect:disks})}, so a collapsing cloud + disk model would be required for a more complete interpretation of these ratios. The origin of the $^{18}$O enrichment in H$_2$CO is less clear, but implies the presence of a latent reservoir of isotopically enriched precursor material, resulting from an unknown fractionation pathway.

\subsection{Isotopic composition of cometary dust and refractories}

Compositional measurements of the non-volatile component of cometary solids (including dust particles, minerals and refractory solids) are much more challenging than for gas measurements, resulting in a paucity of isotopic measurements compared with the cometary ices. The pioneering Stardust mission to comet 81P/Wild 2 in 2006 brought back coma dust samples to Earth for analysis in the laboratory, allowing individual grains (and their sub-components) to be examined in detail, thus revolutionizing our understanding of the composition of cometary refractories. As summarized by \citet{boc15}, the bulk of the material analyzed was surprisingly rich in high-temperature minerals such as crystalline silicates, and conversely, relatively poor in organic material compared with carbonaceous chondrite meteorites. {This may be due to loss of the non-refractory material due to heating during sample collection}. The Stardust samples showed a range of deuterium enrichments from approximately terrestrial up to $\sim3\times$ VSMOW \citep{mck06}. $^{13}$C was found to be slightly depleted ($\delta{^{13}{\rm C}} = -20$ to $-50$ per mil, {where $\delta$X is the fractional enrichment of isotope X relative to a terrestrial standard; see e.g. \citealt{furi15}}). The $^{15}$N/$^{14}$N ratio is similar to the terrestrial value, but with some outliers ($\delta{^{15}{\rm N}} =  +100$ to $+500$ per mil), and in small-scale isotopic `hotspots', the maximum $\delta{^{15}{\rm N}}$ was $\approx+1300$ per mil. Minor depletions were found in $^{18}$O, consistent with chondritic material. The overall dust grain isotopic composition in comet Wild 2 is similar to carbonaceous chondrite meteorites --- \emph{i.e.} significantly enriched in D and N compared with primordial values, and broadly similar to the terrestrial average (albeit with some significant $^{15}$N and D enrichments); see Section \ref{sect:meteorites}.

\citet{pac21} presented the first D/H measurements in the organic refractory component of cometary dust particles, based on in situ analysis by the Rosetta COSIMA instrument at comet 67P. The D/H ratio is an order of magnitude larger than VSMOW, and agrees (within errors) with the heavily deuterium-enriched DCN/HCN value found in comet Hale-Bopp \citep{mei98}. Intermediate between H$_2$O and CH$_3$OH, the moderately-high D/H ratio in carbon-rich cometary dust implies an origin from organic matter synthesized (at least, in part), in the cold, dense ISM. More observations of deuterated organic molecules in comets, and at various stages in the chemical evolution from interstellar cloud to protoplanetary disk, will be crucial for understanding the complex history of refractory carbon in cometary dust samples.

\subsection{Towards a complete understanding of cometary isotopic signatures}

Based on the isotopic ratios of hydrogen, carbon, nitrogen and oxygen, a reasonably consistent picture emerges regarding the origins of cometary ices in cold ($\sim10$--30 K), dense, interstellar/protostellar gas, as a combined result of gas phase and grain-surface chemistry. While some reprocessing of interstellar ices is expected in the protosolar accretion disk (and during the passage of interstellar matter into the disk), strong deuterium enrichment in the disk may be inefficient \citep{cleeves14,fur17}. The physical, chemical and dynamical complexity of the comet-forming environment (see \citealt{est16}, \citealt{eis19} and \citealt{don15}, respectively), together with incomplete information on the state of the protosolar disk at the time of planetesimal formation precludes exact modeling of the cometary accretion process. Nevertheless, a reasonable (at least qualitative) understanding has been achieved regarding the origin of the majority of the observed cometary isotope ratios. 


\section{\textbf{METEORITES/ASTEROIDS}}
\label{sect:meteorites}

\subsection{Background}

The chondritic meteorites are all formed from three basic components: chondrules, refractory inclusions (such as Ca-Al-rich inclusions or CAIs), and matrix \citep{RN7147}. Chondrules and inclusions formed at high temperatures ($\sim$1500-2100 K) in the solar nebula. The fine-grained matrix cements the meteorites and is a mixture of materials that were thermally processed in the solar nebula (e.g., crystalline silicates), as well as lesser amounts of materials like organic matter and presolar circumstellar grains that were not. Based on their chemical, isotopic and physical properties, the chondritic meteorites have been divided into four classes (ordinary, carbonaceous, enstatite and Rumuruti), and subdivided into a number of groups (ordinary - H, L, LL; carbonaceous – CI, CM, CV, CO, CK, CR, CB, CH; enstatite – EH, EL). Each chondrite group is assumed to have come from a separate asteroidal parent body, although it is possible that there are multiple parent bodies with very similar compositions/properties.

The chondrites formed between $\sim$2 Ma and $\sim$4 Ma after CAIs (the oldest dated solar system objects). They accreted as unconsolidated ‘sediments’, but lithification processes in their asteroidal parent bodies produced rocks that were strong enough to survive impact excavation and atmospheric entry. This lithification was driven by internal heating due to the decay of $^{26}$Al (t$_{1/2}\sim$0.7 Ma). Those asteroids that formed at $\sim$2 Ma (ordinary, Rumuruti, enstatite, CK) did not quite get hot enough in their centers to start melting, while heating near their surfaces was relatively mild. For these meteorites, their lithification was predominantly driven by heat (thermal metamorphism) and the extent of their modification will have varied with their depth in their parent bodies. It should be emphasized that, except for the enstatite chondrites, there is clear evidence that they accreted ices, but most experienced high enough peak temperatures to ultimately drive off any water. Parent bodies that formed between 3-4 Ma (e.g., CI, CM, CR) would have had enough $^{26}$Al to melt water-ice and drive reactions that generated hydrous minerals, carbonates, and Fe,Ni-oxides, but generally probably did not reach temperatures above $\sim 420$ K ($\sim$150 $^\circ$C) and can exhibit a wide range in extents of alteration. If planetesimals formed after $\sim$4 Ma, there would not have been enough $^{26}$Al to even melt ice, although impacts could potentially accomplish this at least briefly and locally. Thus, these bodies are unlikely to be the sources of meteorites, but could be sources of some interplanetary dust.

Recently, it has become apparent that in the early solar system the asteroid belt was probably a dynamical dumping ground for planetesimals that formed over a wide range of orbital distances from the Sun and were scattered there by the giant planets. Chondrites{, as well as iron and achondrite meteorites from earlier formed differentiated planetesimals,} are now often classified as carbonaceous and non-carbonaceous (the latter including Earth, Moon and Mars) based on small but systematic isotopic variations in multiple elements (e.g., Ca, Ti, Cr, Mo, Ru) that are nucleosynthetic in origin \citep{RN4486}. The association of the non-carbonaceous group with Earth and Mars, as well as the need to keep the carbonaceous and non-carbonaceous groups physically separated, has led to the suggestion that the non-carbonaceous group formed inside of Jupiter’s orbit and the carbonaceous group outside its orbit \citep{RN6896, RN7232, RN4486}.

\subsection{Carbon and Nitrogen isotopes in meteoritic organics}

In primitive chondritic meteorites, C and N are primarily carried by organic material, with lesser amounts of C in carbonate minerals. On the other hand, H is found predominantly in hydrous minerals (clays, etc.) as OH and H$_{2}$O, with only roughly 10 \% in organic material. Oxygen is distributed between the anhydrous silicates/oxides that were accreted from the nebula, the OH/H$_{2}$O in hydrous silicates, and products of Fe,Ni-metal and -sulfide oxidation by H$_{2}$O. In more thermally metamorphosed samples, dehydration and mineral reactions have erased most signs of any early aqueous alteration, but its influence will still be recorded in their bulk O isotopes and FeO contents. The organics, on the other hand, were destroyed by metamorphism. In most chondrites, the organics were oxidized and lost from all but the least heated samples. In the highly reduced enstatite chondrites, the organics became increasingly graphitized with the loss of H, N and O. Some N may be retained in minerals like TiN, Si$_{3}$N$_{4}$, and Si$_{2}$N$_{2}$O, although there is debate about whether these minerals are metamorphic or formed in the solar nebula. The extent to which aqueous alteration altered organics is even more controversial.

The C isotopic compositions of bulk chondrites fall in a relatively restricted range of $^{12}$C/$^{13}$C $\approx$ 87.5-91.0, the variations largely reflecting differing relative abundances of the more abundant ($^{12}$C$/^{13}$C$\approx$89-92) organic material \cite{RN3161}, and the less abundant carbonate ($^{12}$C$/^{13}$C $\approx$ 82-90) \citep{RN5962, RN6028, RN2241, RN7922, RN7169}. The origin of the carbonate C has yet to be established, but may have been volatile carbonaceous species (e.g., CO$_{2}$, CO, CH$_{4}$) trapped in the accreted ices. In bulk the carbonate C isotopic compositions are quite variable even {within} a group (e.g., $^{12}$C/$^{13}$C $\approx$ 82.9-88.7 in CMs), and vary considerably more from grain to grain in a meteorite. These variations suggest that the carbonate C isotopic compositions probably reflect the evolving conditions (i.e., temperature and gas/fluid compositions) during precipitation more than the source composition(s).

The bulk of the C and N in chondrites is present in organic material, most of which cannot be extracted with solvents \citep{RN7117}. This insoluble material is distributed throughout the matrix in mostly submicron grains that show no obvious spatial relationship to any minerals. Larger grains seem to be aggregates of smaller grains that have collected in veins, perhaps as the result of fluid flow. For reasons that are unclear but may be related to their small grain size, efficient isolation of this insoluble material from chondrites has proved to be extremely difficult, with yields that are typically $\sim$50 $\%$. What is isolated is usually referred to as insoluble organic material (IOM). In terms of H/C ratio, aromaticity and isotopic composition (all indicators of extent of thermal/hydrothermal processing), the most primitive IOM is found in the CR chondrites, the youngest chondrites that probably saw the lowest peak alteration temperatures. The CR IOM has a bulk elemental composition, relative to 100 Cs, of C$_{100}$H$_{75-79}$O$_{11-17}$N$_{3-4}$S$_{1-3}$ \citep{RN3161}, which is not very different to the composition of Halley CHON particles (C$_{100}$H$_{80}$O$_{20}$N$_{4}$S$_{2}$ \citep{RN2077}) and the refractory carbonaceous dust in comet 67P/Churyumov-Gerasimenko (C$_{100}$H$_{104}$N$_{3.5}$ \citep{RN7441, RN7889}, for which O and S were not measured). The CRs also generally have the most isotopically anomalous IOM with bulk D/H $\approx$ 5.5-6.4$\times$10$^{-4}$ and $^{14}$N/$^{15}$N $\approx$ 221-237, although a few ungrouped chondrites are more anomalous with bulk IOM D/H ratios as high as 7.0$\times$10$^{-4}$ and $^{14}$N/$^{15}$N ratios as low as 192. These IOM H isotopic compositions are not as D-rich as the refractory carbonaceous material in 67P (D/H = 1.57$\pm$0.54$\times$10$^{-3}$ \citep{pac21}), but they and the elemental compositions hint at a genetic link, with the material in 67P having experienced less modification in the nebula and/or the comet than the IOM in the chondrites. When analyzed at micron to sub-micron scales, the CR IOM contains a few percent of isotopic hotspots with D/H ratios up to 6.4$\times$10$^{-3}$ and $^{14}$N/$^{15}$N ratios as low as 68. These hotspots seem to be associated with individual grains, but there is no straightforward correlation between the D and $^{15}$N enrichments \citep{RN2856, RN6123}.

IOM has been isolated from at least some members of all chondrite groups except the Rumuruti and CK chondrites, which are too metamorphosed to have preserved much, if any IOM. The IOM in the non-CR chondrite groups is generally not as isotopically anomalous and has lower H/C ratios than the CR IOM that, if all IOM had the same or similar precursors, are an indication of heating. This heating most likely occurred in their parent bodies, but some heating in the disk prior to accretion cannot be ruled out. Interplanetary dust particles also contain significant amounts of refractory, insoluble organic C \citep{RN2365, RN1356}. In the anhydrous chondritic porous (CP-) IDPs, which are likely to have come from comets, most or all the H should be associated with organic material. Measurements of their H isotopic compositions are generally not as anomalous as the IOM in the CRs, for instance, when analyzed at similar scales \citep{RN3700}. However, like the IOM, there are hotspots that can have similarly extreme D and $^{15}$N enrichments. The less anomalous isotopic compositions of CP-IDPs could simply be the result of atmospheric entry heating driving off the more thermally labile and isotopically anomalous functional groups.

Although not as abundant as the IOM, there are very complex suites of solvent-soluble organics in the primitive chondrites \citep{RN7566, RN4076}. The absolute and relative abundances of the different molecules differ from meteorite to meteorite in ways that are still poorly understood, but they seem to be a function of the extent of aqueous alteration. These variations indicate that the aqueous chemistry in the meteorites was quite dynamic, and what is present now is almost certainly not the same as what was accreted \citep{RN8781}. The best studied family of soluble molecules are the amino acids. They also tend to be the most isotopically anomalous, with D/H $\approx$ 1.65-12.6$\times$10$^{-4}$, $^{14}$N/$^{15}$N $\approx$ 205-269, $^{12}$C/$^{13}$C $\approx$ 84.5-91.0 \citep{RN4849, RN3975}, albeit with considerable variability within and between meteorites. The most abundant amino acids in the least altered chondrites are the so-called $\alpha$-amino acids that are thought to have formed by reaction of simpler ketones and aldehydes with NH$_{3}$ and HCN in aqueous solutions (Strecker-cyanohydrine synthesis).

The presence of NH$_{3}$ and HCN in the aqueous solutions again is suggestive of there being a link between the ices accreted by the chondrites and those in comets. The simplest amino acid, glycine (D/H $\approx$ 2.0-3.5$\times$10$^{-4}$, $^{14}$N/$^{15}$N $\approx$ 237-262, $^{12}$C$/^{13}$C $\approx$ 85.0-87.8 \citep{RN4849, RN3975}, would have formed from formaldehyde with one C added from HCN and N added from NH$_{3}$. The $^{12}$C/$^{13}$C ratio of formaldehyde in 67P is 40$\pm$14 \citep{alt20}, which is isotopically much heavier than formaldehyde in the Murchison meteorite ($^{12}$C/$^{13}$C $\approx$ 83.4 \citep{RN7760}), or glycine or any other amino acid measured in chondrites. The $^{14}$N/$^{15}$N ratios of NH$_{3}$ and HCN in comets are very uniform ($\sim$140) and again much heavier than in glycine in chondrites. Unless HCN has a very light $^{12}$C$/^{13}$C ratio to compensate for the heavy composition of the formaldehyde and there are as yet unrecognized sources of $^{15}$N-depleted NH$_{3}$  in comets, the glycine isotopic compositions in chondrites point to differences in the nature of the ices accreted by chondrites and primitive comets like 67P.

\subsection{Hydrogen isotopes in meteoritic water and organics}

While the organic material can be very enriched in D, the D/H of the bulk meteorites are much lower. For instance, for the CMs the IOM D/H $\approx$ 2.5-3.4$\times$10$^{-4}$, while the bulk meteorites have D/H $\approx$ 1.2-1.8$\times$10$^{-4}$. The reason for this difference is that the water accreted by the chondrites was isotopically much lighter than the organics. How much lighter can be estimated by plotting D/H vs. C/H for the bulk meteorites in a group that have significant variation in the extents of alteration. In this case, the bulk compositions should produce a linear trend reflecting variable mixing of the H in water (now in clay minerals, etc.) and organics, with projection to C/H = 0 giving an estimate of the D/H of the water and extension of the trend to high C/H passing through the initial bulk D/H of the organics. In this way, \citet{RN4810} estimated that the water accreted by the CMs and the CIs had a D/H $\approx$ 8.6$\times$10$^{-5}$, and that the initial bulk organic D/H was similar to that of IOM in the CRs (i.e., there has been D/H exchange between water and organics during alteration). The CRs, on the other hand, seem to have accreted more D-rich water with a D/H $\approx$ 1.7$\times$10$^{-4}$.

The D/H of the water accreted by the other carbonaceous chondrite groups is more difficult to estimate because metamorphism has resulted in at least partial dehydration and modification of the organics. The water D/H ratios in CI, CR, CM and CV chondrites have been estimated by measuring in situ the D/H and C/H ratios of multiple areas 10-15 $\mu$m across of bulk meteorites \citep{RN7685, RN5898, RN7449}. The estimated D/H ratios of 1.55$\times$10$^{-4}$ for the CIs, 1.0$\times$10$^{-4}$ for the CMs and 2.38$\times$10$^{-4}$ for the CRs are more D-rich than the estimates based on bulk measurements above. The D/H $\approx$ 1.4$\times$10$^{-4}$ for the CVs is intermediate between the bulk estimates for the CMs/CIs and CRs. If there has been water-organic H isotopic exchange, for instance, the in situ measurements will reflect the current average D/H ratios of the water and organics rather than their initial compositions. However, Piani et al. interpret their results as reflecting the accreted water compositions, so that the variations that they observe amongst the CMs imply heterogeneous accretion of at least two isotopically distinct ices. \citet{RN5898} reported possible evidence for even more dramatic H isotopic heterogeneity in ices accreted by the primitive ordinary chondrite Semarkona, with compositions of 1.5$\times$10$^{-4}$ and 1.8$\times$10$^{-3}$. However, \citet{RN8708} have attributed this range to terrestrial contamination and parent body processes (see below).

There are two sources of uncertainty in the estimates of the water D/H. First, the extent of terrestrial contamination by atmospheric water in the samples analyzed by \citet{RN4810} have probably been underestimated \citep{RN8788, RN8406}, which will change both the C/H and D/H values. How much this will change the estimates of the water D/H remains to be seen, but it is likely to affect the CRs more than the CMs. Terrestrial contamination will also have effected in situ measurements, but the extent to which contamination has compromised them has also not been quantified. Secondly, it is possible that the D/H of the water was modified during the early stages of aqueous alteration when metal was oxidized \citep{RN3832}. Reactions such as 3Fe + 4H$_{2}$O = Fe$_{3}$O$_{4}$ + 4H$_{2}$ would have generated copious amounts of H$_{2}$. So much H$_{2}$, in fact, that unless it found ways to escape to space the pore pressure would have exceeded the tensile strength of an asteroid, leading to its catastrophic disruption. At temperatures of 270-470 K (0-200 $^\circ$C), isotopic exchange between H$_{2}$O and H$_{2}$ is relatively facile \citep{RN7658} and there is a very large isotopic fractionation with the H$_{2}$ being very depleted in D. On Earth, for instance, H$_{2}$ with D/H $\approx$ 3.1-6.2$\times$10$^{-5}$ has been found escaping areas undergoing low temperature serpentinization (aqueous alteration that is quite similar to what meteorite experienced). Continuous loss of such D-poor H$_{2}$ will result in the remaining water becoming increasingly D-rich, with the degree of D enrichment depending on the alteration temperature and the initial water/metal ratio. Based on the valence state of Fe in the bulk meteorites, estimated current water and initial metal contents, and assumed alteration temperatures of 270-470 K (0-200 $^\circ$C), \citet{RN7124} estimated for the CIs the initial water D/H $\approx$ 5.1-9.0$\times$10$^{-5}$, for the CMs D/H $\approx$ 5.1-7.9$\times$10$^{-5}$, and for the CRs D/H $\approx$ 7.4-14.7$\times$10$^{-5}$.

Whether one uses the initial estimates or the valence corrected ones, most or all of the CI-CM-CR water D/H ratios fall between the terrestrial and solar ratios. These are significantly lower than in almost all measured comets. Perhaps the simplest explanation for the range of water D/H ratios in comets and carbonaceous chondrites is that they reflect variable mixtures of interstellar water and water that re-equilibrated with H$_{2}$ at high temperatures in the disk \citep{cleeves16, RN4625, RN5095}. If correct, then the carbonaceous chondrites accreted small but significant amounts of interstellar ices, which would be qualitatively consistent with the presence of presolar circumstellar grains in these chondrites and also might point to an ultimately interstellar origin for their organic matter \citep{RN7221}.

The H isotopic fractionation associated with metal oxidation is likely to have had an even bigger effect in the ordinary and Rumuruti chondrites in which water/metal ratios were much lower than in the three carbonaceous groups. In Semarkona (LL3.0), the least metamorphosed ordinary chondrite, the bulk water D/H is estimated to be 2.8-3.4$\times$10$^{-4}$ \citep{RN4810}, and hydrous minerals in one Rumurutiite have D/H $\approx$ 7.3$\pm$0.1$\times$10$^{-4}$ \citep{RN3548}. If the D enrichments are the result of parent body processes, \citet{RN7124} estimated that Semarkona’s initial water D/H $\approx$ 7.4-18.0$\times$10$^{-5}$, which overlaps, within the large uncertainties, with the initial water D/H estimates for the carbonaceous chondrites. If the high water D/H ratios in ordinary and Rumuruti chondrites, which are comparable to the most D-rich comets, are not the result of parent body process, they accreted a higher fraction of interstellar water than comets and carbonaceous chondrites. This would be surprising given that the ordinary and Rumuruti parent bodies formed in the inner solar system relatively early when the ambient nebula temperatures are expected to have been higher than in the formation regions of comets and carbonaceous chondrites in the outer solar system.

\subsection{Oxygen isotopes in meteorites}
\label{meteorites-oxygen}

To first order, the range of O isotopic compositions in solar system materials reflect variations in $^{16}$O relative to $^{17,18}$O (Figure~7). These so-called mass independent isotopic variations are quite different to the mass dependent isotopic variations that are produced by typical physical and chemical processes in which the relative change in the $^{16}$O$/^{17}$O ratio is roughly half that of the $^{16}$O$/^{18}$O ratio. Another striking feature of most solar system materials, is that their O isotopic compositions fall relatively close to that of the Earth ($^{16}$O$/^{18}$O=498.7, $^{16}$O$/^{17}$O=2732) and are significantly more $^{16}$O-poor than the solar O isotopic composition ($^{16}$O$/^{18}$O=530, $^{16}$O$/^{17}$O=2798) \citep{RN8814, RN4400, RN6908}. This is true even for CP-IDPs that may come from comets \citep{RN4572, RN5747, RN4961} and crystalline silicates returned from comet Wild 2 \citep{mck06, RN3528, RN5741}. The only objects whose O isotopic compositions approach those of the solar composition are refractory inclusions \citep{RN8655, RN6908}, which are rare in Wild 2 dust and CP-IDPs. Only one object has been found with an O isotopic composition ($^{16}$O$/^{18}$O=539, $^{16}$O$/^{17}$O=2846) that is more $^{16}$O-rich than solar \citep{RN5838}.

\begin{figure}[h!]
\centering
\includegraphics[width=\columnwidth]{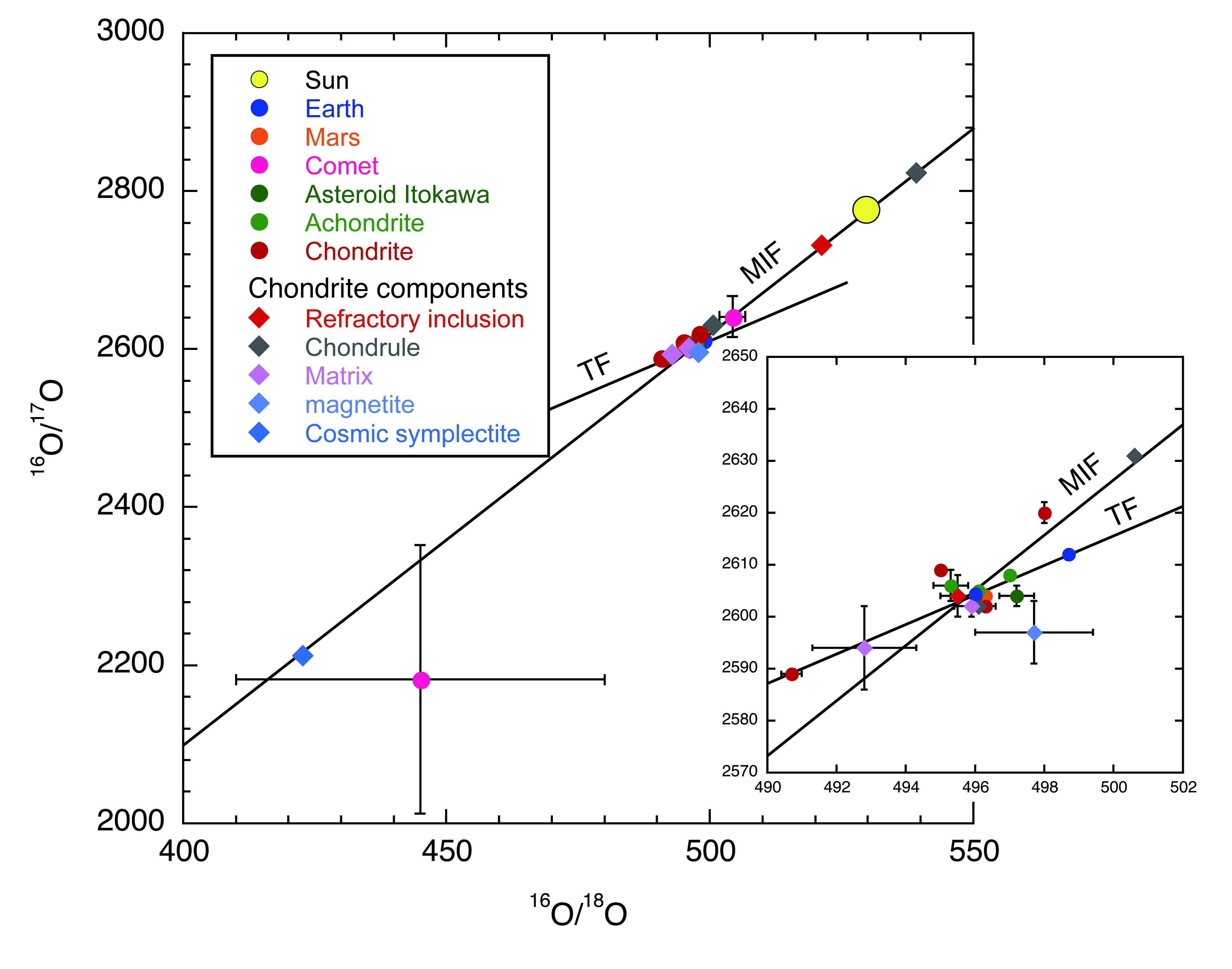}
\caption{The variations in $^{16}$O/$^{17}$O and $^{16}$O/$^{18}$O ratios amongst selected solar system materials. TF is the terrestrial mass fractionation line. MIF is the mass independent fractionation line resulting from the fractionation of $^{16}$O from $^{17}$O and $^{18}$O.}
\end{figure}

The generally accepted mechanism for producing the mass independent O isotopic variations is by self-shielding during UV dissociation of CO (Section 2.3) that produces $^{17,18}$O-rich water and $^{16}$O-rich CO. To date, the most extreme $^{17,18}$O-enrichment that has been found in any meteoritic material ($^{16}$O$/^{18}$O=423, $^{16}$O$/^{17}$O=2231) is in so-called cosmic sympectite \citep{RN3860} and provides some idea of what the O isotopic composition of the water generated by self-shield might have been. The $^{16}$O$/^{18}$O ratio of the cosmic sympectite is similar to that of the water in comet 67P (Section 4.4). Because of its volatility and the fact that it is relatively unreactive, to date it has not been possible to directly constrain the O isotopic composition of the CO.

It has been proposed that the CO UV self-shielding took place either in the protosolar molecular cloud \citep{yurimoto04} or the outer solar system \citep{lyons05}, although several arguments seem to favor a molecular cloud origin \citep{RN7221, RN8655}. Transferring the $^{17,18}$O-enrichments in water to silicates requires high temperatures and separation of the water-ice from the CO. This and the fact that most solar system materials have O isotopic compositions that are fairly close to those of the Earth and Mars, both \citet{yurimoto04} and \citet{lyons05} proposed that there was a massive influx of $^{17,18}$O-rich ice from the outer solar system into the inner solar system that changed its bulk composition from solar-like to terrestrial-like. If the carbonaceous chondrites formed beyond the orbit of Jupiter, the O isotopic compositions of most of their components indicates that if such an influx occurred it must have affected some regions of the outer solar system as well. Also, if all silicates with non-solar, more terrestrial-like O isotopic compositions formed inward or somewhat beyond the orbit of Jupiter, this would imply considerable outward radial transport of a large fraction of the material in comets like Wild 2 and the possibly cometary parent bodies of the CP-IDPs.

However, there are two potential mechanisms for producing terrestrial-like O isotopic compositions in silicates anywhere in the solar system if heat sources can be found (e.g., FU Orionis events, impacts and shocks). One mechanism is to simply concentrate silicate dust and ice relative to CO in the gas \citep{RN7221}. In this case, the O isotopic composition rapidly asymptotes to the bulk composition of the silicates plus ice, which is terrestrial-like, with increasing (dust+ice)/gas ratio. This scenario would also predict that refractory inclusions formed in regions with very modest (dust+ice)/gas enrichments, relative to solar, a conclusion that others have reached based on the valence states of Ti and V \citep{RN6580}. It also predicts that objects that have O isotopic compositions that are more $^{16}$O-rich than solar should be rare because they would only form in (dust+ice) depleted regions. An alternative mechanism is suggested by experiments that show that amorphous forsterite and enstatite exchange O isotopes with water vapor at relatively low temperatures (e.g., complete isotopic exchange if heated to 500-650 K for the lifetime of the disk), but the less reactive CO does not \citep{RN7711, RN8815}. The attraction of this mechanism is that most or all interstellar silicates are amorphous and one would not need to even concentrate the silicate dust and ice relative to the gas. 

\section{\textbf{PLANETARY ATMOSPHERES}} \label{sec:planets}

In this section we discuss the complexities in attempting to explain present-day isotopic ratios in planetary atmospheres, the regimes for which we have the most data points across the solar system (meteoritic ratios are discussed in $\S \ref{sect:meteorites}$). There are two important considerations: (1)	The value at time of formation; (2) subsequent enrichment/de-enrichment processes.

For (1) --- formation --- planets form a middle ground between the protoplanetary disks discussed in \S3, and the small icy bodies such as comets discussed in \S \ref{sect:comets}. In essence, planets may accrete from both gaseous and solid reservoirs with different molecular constituents and different inherent isotopic ratios due to previous fractionation processes. For example, it is well established that hydrogen exchange occurs leading to D/H in \ce{H2} (gas phase) being typically much lower than D/H in ices such as \ce{CH4}, \ce{H2CO} and \ce{NH3} (Fig.~\ref{fig:ratios}) \citep{mousis02}. Since different planets accrete different proportions of ice and gas depending on their mass and distance from the star in the proto-planetary disk (e.g. compare Earth to Jupiter) the original bulk isotopic ratios are thus also different. This is the first cause of isotopic ratio variations, and most evident for D/H in planetary atmospheres.

In the case of (2) --- processes --- planetary atmospheres host many different isotope-selective physical and chemical phenomena, including some not possible for disks or molecular clouds. Important early work on isotopic evolution of planetary atmospheres was performed by Kasting, Hunten and colleagues \citep{mcelroy69, hunten73, hunten82, kasting83, hunten87, pepin91}. It is now known that many processes act to change atmospheric isotopic ratios, including:

\begin{enumerate}
\item	Exchange processes \citep{thiemens99};
\item	Isotope-selective photo-dissociation \citep{Liang2007};
\item	Kinetic Isotope Effect (KIE) in chemical reactions, including gas phase \citep{pinto86, Nixon2012} and solid formation \citep{sebree16};
\item	Non-thermal escape processes such as ion pick-up \citep{lammer20}, {where molecules are first ionized and then become attached to the magnetized plasma of the solar wind and escape;} and photochemical production of suprathermal ions \citep{jakosky94};
\item	Other mass-dependent escape processes, including thermal (Jeans) escape \citep{hunten73}, sputtering (Johnson 1990), and possible early hydrodynamic escape \citep{hunten87};
\item	Other non-escape effects that may fractionate molecules or element onto the surface or into interior, for example condensation \citep{montmessin05}, dissolution in seas and oceans \citep{benson84}, enclathratization in ice matrix \citep{hesse81}, and others.
\end{enumerate}


It is beyond the scope of this review to describe all of these processes in detail – see review by \citet{lammer18} and references therein.  These fractionation processes lead to further variations in D/H and also in \nratio\ between the atmospheres of terrestrial planets, although apparently much less in the case of \cratio\ and \oratio.

In the following sections we will review the current understanding of isotopic ratios in H, C, N and O in the planetary atmospheres of the solar system.

\subsection{Hydrogen isotopes}

Hydrogen D/H is the most variable isotopic ratio across the solar system, from $\sim2 \times 10^{-5}$ on the giant plants \citep{pierel17}, 
$\sim4{\times}10^{-5}$ on the ice giants \citep{feuchtgruber13}, to 
$\sim1.5{\times}10^{-4}$ on Titan and Earth \citep{niemann10, hagemann70}, $9{\times}10^{-4}$ on Mars \citep{webster13} and $1.6{\times}10^{-2}$ on Venus \citep{donahue82}. 

This variation is explained by two effects. The first sets a primordial difference due to differing gas (\ce{H2}) to ice (\ce{CH4}, \ce{H2O}, \ce{NH3}) ratios, with the gas giants (Jupiter and Saturn) having the highest fractions of gas, the terrestrial planets the lowest values closest to ices, and the ice giants (Neptune, Uranus) having intermediate values. For terrestrial planet atmospheres, differing rates of water photolysis and escape of H \emph{vs.} D, have led to increasing D/H from Earth (least fractionated) to Mars to Venus (most fractionated) {Therefore, it is important to note that the processed isotopic ratios found in these atmospheres may not reflect the primordial or current bulk ratios in the interior. }

D/H values on Titan and Enceladus \citep{niemann10,waite09} appear within a factor 2 of terrestrial D/H, and within the cometary range \citep{boc15}, reflecting an icy composition. There is some evidence for fractionation between gases on Titan (\ce{CH4}, \ce{C2H2}, HCN and \ce{H2}) \citep{Nixon2012, coustenis08, Molter2016, niemann10}, however error bars are substantial at this stage, and further work is required to definitively say whether for example there are kinetic isotope effects in play. Some results \citep{Lellouch01, pierel17} have indicated that the D/H on Saturn may be slightly lower than on Jupiter, contrary to the expected trend. This result is surprising and needs further work to reduce error bars and confirm.

In addition to long-term evolution, D/H in Martian atmospheric water has been found to exhibit spatial and temporal variation due to seasonal effects, as revealed by measurements for from ground-based and spacecraft observations \citep{villanueva15, alday21}.

\subsection{Carbon isotopes}

Mars has the only variation of \cratio\ that is demonstrably incompatible with a terrestrial value. The $^{13}$C/$^{12}$C enrichment of  $+4.6 \pm 0.4$ \% in the atmosphere {(vs terrestrial PBD)} \citep{webster13} may be compared with  $-2.5 \pm 0.5$ \% in the mantle \citep{wright92}. This divergence in atmospheric versus geological \cratio\ {can} be explained if the early Martian atmosphere was CO and \ce{CH4} with little \ce{CO2}. CO and \ce{CH4} are depleted in $^{13}$C and escape, while $^{13}$C-rich CO$_2$ is incorporated into the surface \citep{galimov00}, mostly prior to 3.9 Gya. In addition, CO may preferentially lose $^{13}$C due to less self-shielding \citep{hu15}. \ce{CO2} may also fractionate due to preferential dissociation of $^{12}$\ce{CO2} \citep{schmidt13} and so leave $^{13}$C-rich \ce{CO2} to be incorporated into surface carbonates.

Recent work on carbon isotopes in multiple species on Titan from Cassini and ALMA have shown no strong evidence for fractionation of \cratio\ from methane to daughter species such as hydrocarbons, nitriles and CO$_x$ species in the atmosphere \citep{nixon08a, Nixon2012, Molter2016, Serigano2016, Jennings2008, nixon08b, jolly10, iino21}, although error bars are currently larger than in martian measurements so it may be too soon to definitively rule out any fractionation.

\subsection{Nitrogen isotopes}

Nitrogen, in contrast to neighboring elements on the periodic table carbon and oxygen, exhibits strong variations in isotopic ratio throughout the solar system. This indicates that planetary differences in light element isotopic ratios (other than H) were not primordial, since in that case C and O would be expected to show strong differences between bodies, similar to N. Rather, we conclude that N heterogeneity developed later due to fractionation.

\citet{furi15} have argued that there are three distinct reservoirs of nitrogen evident today: the high \nratio\ solar and solar-like ratios seen in Jupiter and Saturn of $\sim$500; the lower values $\sim$270 seen in the inner solar system on Venus,  Earth and Moon, and Martian rocks; and finally a still-lower reservoir of $<$200 seen on Titan and in comets (see Fig.~\ref{fig:ratios}).

An outlier is the atmosphere of Mars, with a ratio of $\sim$170 \citep{wong13}, much lower than recorded in Martian rocks \citep{mohapatra03}. This indicates substantial escape of isotopically light nitrogen early in its history, which must have occurred early in the solar system when solar EUV was higher to cause hydrodynamic escape, due to efficient cooling by \ce{CO2} at the present day \citep{lammer18}. This difference from Earth and Venus may be due to Mars’ lower mass and gravity.

Finally, on Titan an enrichment of $\sim$3 is seen in \nratio\ from the bulk reservoir of \ce{N2}, to the photochemical products (nitriles) \citep{Marten2002, Cordiner2018, Vinatier2007, Molter2016, Iino2020}. An explanation has been offered in terms of self-shielding by ($^{14}$N)$_2$ against photolysis, while the less abundant $^{14}$N$^{15}$N has no such shielding and therefore photolyzed at a higher rate, providing an atomic source of N which is enriched in $^{15}$N, and subsequently incorporated into CN and other nitrogen-bearing molecules \citep{Liang2007}. The recent Titan photochemical model of \citet{vui19} incorporates updated molecular nitrogen photolysis cross sections, and provides reasonably good agreement with the level of $^{15}$N-enrichment observed by ALMA in HCN, HC$_3$N and CH$_3$CN. However, there appears to be a tendency for the model to overestimate the $^{15}$N enrichment in HCN and CH$_3$CN, which may be a result of incomplete understanding of Titan's complex atmospheric nitrogen chemistry.

\subsection{Oxygen isotopes}

Oxygen \oratio\ does not show strong variations in planetary atmospheres, at the levels seen in nitrogen or hydrogen (see Fig.~1). Smaller variations however have been noted in Earth minerals and meteoritic samples, discussed further in \S \ref{meteorites-oxygen}. Further observational astronomy and/or probes are needed to determine the \oratio\ on the giant planets, since measurements are currently lacking for all four of these bodies other than Jupiter, for which the ratio has a factor of two in error bar \citep{noll95}.






\section{\textbf{SUMMARY}}

Recent high sensitivity astronomical observations, robotic solar system exploration missions, and detailed analysis of meteorites have made it possible to greatly increase the isotopic measurements in star and planet forming regions as well as solar system objects. Meanwhile, recent developments in chemical modeling have contributed to knowledge of the  isotopic links from planet forming regions to the solar system.

\begin{itemize}
\item Deuterated species are the most well-studied isotopologues, particularly in the context of the origin of water on the Earth. Recent high spatial resolution observations of protoplanetary disks by ALMA together with chemical modelling suggest warm paths of molecular deuteration in the inner parts of disks, in addition to well-known cold paths in the outer disk.

\item Deuterium ratios of various molecular species measured in the comet 67P by Rosetta have provided us with new insights. Models have elucidated the formation processes and epochs of the observed molecules. Isotopic ratios in refractory grains in 67P have also been measured, which show a higher deuterium ratio than the primitive IOM found in CR chondrites, despite their similarity in bulk elemental composition.

\item Deuterium ratios in water accreted by various types of meteorites are estimated by measuring the decrease of the deuterium ratios in organics due to aqueous {alteration}, which show significantly lower values than those in comets and even the Earth's ocean. These estimations suggest a decrease of deuterium fractionation of interstellar water at high temperature in the protosolar nebula.

\item Deuterium ratios on Titan and Enceladus are similar to those in comets, suggesting an icy origin, while some possible fractionation in different species have been measured with still significant error bars in the Titan atmosphere.

\item Oxygen isotopic ratios in meteorites are interpreted to link to isotope-selective photodissociation of CO and formation of heavy water ice in the star and planet forming region. Oxygen isotope ratios measured in various molecules in the comet 67P are consistent with this scenario and theory of molecular formation in the interstellar medium.

\item Transfer of oxygen isotopes from water ice to silicate in meteorites is related to the conditions of the protosolar nebula, such as high temperatures, dust+ice-to-gas ratios, and crystallinity (amorphous vs. crystalline) of silicates.

\item Nitrogen fractionation path through isotope selective photodissociation of \ce{N2} is proposed, which is qualitatively consistent with recent observations in prestellar/protostellar cores and protoplanetary disks, depending on the amount of small dust grains. In order to understand the N fractionation measurements in \ce{N2} in the comet 67P in this context, further investigation on the origin of \ce{N2} is needed.

\item Nitrogen isotopic ratios have some variation in the solar system objects, partly due to local processes. ALMA observations of nitrogen fractionation in X-CN organics in Titan’s atmosphere can be explained by photochemistry, as is the case of irradiated, dense molecular clouds.

\item Carbon isotopic ratios are known to be relatively flat among the solar system objects. \cratio\ in various molecular species are measured in the comet 67P by Rosetta, and only \ce{H2CO} shows strong fractionation, which resembles observations in a high-mass protostellar core, though further confirmation of such an association will be required. Amino acids in meteorites are known to be isotopically anomalous, but the fractionation is generally less significant compared with those found in comets. 

\end{itemize}

There is still a long way to go to fully understand how material evolves from the planet forming region to solar system objects, although there have been significant new findings observationally and theoretically. Future observations of cold and hot gas-phase molecular isotopologues as well as molecular ices in star and planet forming regions will contribute greatly to our understanding. For example, deeper molecular line surveys using ALMA and other facilities, and high dispersion infrared spectroscopy using thirty meter-class large telescopes, together with future solar system exploration missions will give us further insights on the isotopic link. Theoretically, {how to treat isotope chemistry of both gas and ice 
is the key to understanding the link from astronomical observations of gas in the planet forming regions and primordial solid materials in the solar system. Understanding} chemical evolution together with dynamical evolution of gas and dust from prestellar/protostellar cores, protosolar nebula to formation of planetesimals and then planets will provide us clues to better understand the origins of solar system objects.
{Our knowledge of the isotopic links from planet forming regions to the solar system could be potentially expanded to derive information on the origin of exoplanetary systems from future observations of isotope ratios in exoplanetary atmospheres 
in future.}

\bigskip

\noindent\textbf{Acknowledgments} 
We acknowledge P. Hily-Blant, K. Altwegg, Y. Kebukawa, M. Hashiguchi, and D. Yamamoto for their valuable inputs to our chapter, and G. Cataldi and Y. Yamato for kindly providing us data of a figure.

\bibliography{pp7}

\end{document}